\documentclass[aps,prb,twocolumn,groupedaddress,showpacs,floatfix,nofootinbib]{revtex4}
\usepackage{graphics}
\usepackage{epsfig}
\usepackage{subfigure}
\usepackage{multirow}

\newcommand{\angs}{\mathrm{\AA}}
\newcommand{\magneton}{\mu_\mathrm{B}}

\newcommand{\be}{\begin{equation}}
\newcommand{\ee}{\end{equation}}

\newcommand{\bea}{\begin{eqnarray}}
\newcommand{\eea}{\end{eqnarray}}

\newcommand{\eqn}[1]{Eq.~(\ref{#1})}
\newcommand{\eqns}[2]{Eqns.~(\ref{#1}) and~(\ref{#2})}

\newcommand{\eqnsfour}[4]{Eqns.~(\ref{#1}),~(\ref{#2}),~(\ref{#3}) and~(\ref{#4})}

\newcommand{\reffig}[1]{Fig.~\ref{#1}} 
\newcommand{\reffigs}[2]{Figs.~\ref{#1} and~\ref{#2}}
\newcommand{\reffigss}[3]{Figs.~\ref{#1},~\ref{#2} and~\ref{#3}}
\newcommand{\reftab}[1]{Table~\ref{#1}}
\newcommand{\reftabs}[2]{Tables~\ref{#1} and~\ref{#2}}

\newcommand{\eform}{E_\mathrm{f}}
\newcommand{\emig}{E_\mathrm{m}}
\newcommand{\ebind}{E_\mathrm{b}}

\begin{document}

\title{First principles study of helium, carbon and nitrogen in
  austenite, dilute austenitic iron alloys and nickel.}

\author{D.J.~Hepburn}
\email[Email: ]{dhepburn@ph.ed.ac.uk}
\author{D. Ferguson}
\author{S. Gardner}
\author{G.J.~Ackland}
\email[Email: ]{gjackland@ed.ac.uk}
\affiliation{Institute for Condensed Matter and Complex Systems, School of Physics and SUPA, The University of Edinburgh, Mayfield Road, Edinburgh, EH9 3JZ, UK.} 
\date{\today}
\pacs{61.72.-y,61.82.Bg,71.15.Mb,75.50.Bb,75.50.Cc,75.50.Ee}

\begin{abstract}

An extensive set of first-principles density functional theory
calculations have been performed to study the behaviour of He, C and N
solutes in austenite, dilute Fe-Cr-Ni austenitic alloys and Ni in
order to investigate their influence on the microstructural evolution
of austenitic steel alloys under irradiation. The results show that
austenite behaves much like other fcc metals and like Ni in
particular. Strong similarities were also observed between austenite
and ferrite. We find that interstitial He is most stable in the
tetrahedral site and migrates with a low barrier energy of between 0.1
and 0.2 eV. It binds strongly into clusters as well as overcoordinated
lattice defects and forms highly stable He-vacancy (V$_m$He$_n$)
clusters. Interstitial He clusters of sufficient size were shown to be
unstable to self-interstitial emission and VHe$_n$ cluster
formation. The binding of additional He and V to existing V$_m$He$_n$
clusters increases with cluster size, leading to unbounded growth and
He bubble formation. Clusters with $n/m$ around 1.3 were found to be
most stable with a dissociation energy of 2.8 eV for He and V
release. Substitutional He migrates via the dissociative mechanism in
a thermal vacancy population but can migrate via the vacancy mechanism
in irradiated environments as a stable V$_2$He complex. Both C and N
are most stable octahedrally and exhibit migration energies in the
range from 1.3 to 1.6 eV. Interactions between pairs of these solutes
are either repulsive or negligible. A vacancy can stably bind up to
two C or N atoms with binding energies per solute atom up to 0.4 eV
for C and up to 0.6 eV for N. Calculations in Ni, however, show that
this may not result in vacancy trapping as VC and VN complexes can
migrate cooperatively with barrier energies comparable to the isolated
vacancy. This should also lead to enhanced C and N mobility in
irradiated materials and may result in solute segregation to defect
sinks. Binding to larger vacancy clusters is most stable near their
surface and increases with cluster size. A binding energy of 0.1 eV
was observed for both C and N to a [001] self-interstitial dumbbell
and is likely to increase with cluster size. On this basis, we would
expect that, once mobile, Cottrell atmospheres of C and N will develop
around dislocations and grain boundaries in austenitic steel alloys.

\end{abstract} 

\maketitle

\section{Introduction}

Steel, in its many forms, is the primary structural material in
current fission and fusion systems and will be so for the foreseeable
future. Carbon (C) and nitrogen (N) are both commonly found in steel,
either as important minor alloying elements or as low concentration
impurities. In body centred cubic (bcc) $\alpha$-iron ($\alpha$-Fe),
it has been shown experimentally that C interacts strongly with
vacancy point defects and more weakly with self-interstitial
defects\cite{Vehanen82,Takaki83} and can form so called Cottrell
atmospheres around dislocations\cite{Cottrell49}, influencing yield
properties and leading to strain ageing of the material. First
principles (ab initio) calculations, as summarised in a recent review
by Becquart and Domain\cite{Becquart12}, support these findings and
demonstrate that N exhibits similarly strong interactions. As such,
both of these elements have a significant influence on microstructural
evolution in bcc Fe, even down to very low concentrations, and a
detailed understanding of their interactions and dynamics in steels is
worthy of development, more generally.

Helium (He) is produced in significant quantities in the high
neutron-irradiation fluxes typically experienced by the internal
components of fission reactors and in the structural materials for
fusion systems by (n,$\alpha$) transmutation reactions. In combination
with the primary point defect damage typical of irradiated
environments, the presence of He plays a critical role in the
microstructural evolution of these materials. As a result of its low
solubility in metals, He becomes trapped in regions of excess volume,
such as dislocations, grain boundaries and, most strongly, in
vacancies and vacancy
clusters\cite{Edwards74,Philipps82,Sciani83,Vassen91,Fedorov9698,vanVeen03,Cao11,Ono11}. As
such, it aids the nucleation, stabilisation and growth of voids (He
bubbles), resulting in swelling of the
material\cite{Mansur86,Murase98,vanVeen03,Liu04,Lei11}. The formation
of He bubbles has also been implicated in high-temperature
embrittlement of materials\cite{Schroeder85,Katoh03,vanVeen03}. It is
therefore of critical importance to gain a deep understanding of the
behaviour of He in these materials and the part it plays in the
underlying mechanisms of microstructural evolution.

First principles electronic structure calculations offer the most
accurate means to develop an atomic level understanding of the
dynamics and interactions of solutes and point defects in solids. As
such, they play a central role in the development of a theoretical
understanding of the microstructural evolution of irradiated
materials, as part of a multi-scale modelling approach, such as that
used in the FP6 project, PERFECT\cite{Perfect} and the FP7 project,
PERFORM60\cite{Perform60}.

The behaviour and interactions of He in a number of bcc and
face-centred cubic (fcc) metals have been studied using density
functional theory (DFT)
techniques\cite{Perfect,Zu,DomainPerfect,Yang,Zeng09,Fu0507,SeletskaiaA,Becquart12,Becquart06,Becquart09,Becquart10,Zhang11}. This
database of He kinetics and interactions is essential for the
interpretation of complex experimental results, such as those present
in thermal He desorption spectra. A case in point is the work of Ortiz
{\it et al.}\cite{Ortiz0709}, who have developed a rate theory model
based on DFT calculations of the kinetics and interactions of point
defects, He and C in bcc Fe\cite{Fu0507,Fu05NMat,Ortiz0709}. The model
successfully reproduces and interprets the existing experimental
desorption results\cite{Vassen91}. It is interesting to note that
agreement with experiment was only possible once the effects of C were
included, even though only 150 at. ppm of C was necessary; this again
indicates the sensitivity of the microstructural evolution to C
concentration. To date, however, there have been no ab initio studies
of He in austenite, that is fcc $\gamma$-Fe, or austenitic FeCrNi
alloys. This, primarily, is a result of the difficulty in describing
the paramagnetic state of these materials.

Ab initio calculations have also been used to extensively study
C\cite{JiangCarter,Domain04,Forst06,Lau07,Fu08,Ortiz0709,Becquart12}
and N\cite{Domain04,Becquart12} in bcc Fe. These calculations show
excellent agreement with experimentally verifiable parameters, such as
the migration energy barrier for C diffusion, where ab initio values
of 0.86 eV\cite{JiangCarter,Lau07}, 0.87 eV\cite{Fu08} and 0.90
eV\cite{Domain04} are in good agreement with the experimental value of
0.87 eV\cite{Wert50,Takaki83}. For N, an equally good agreement is
seen for the migration barrier, where a value of 0.76 eV was found by
ab initio calculations\cite{Domain04} and a value of 0.78 eV was found
experimentally\cite{Thomas54}. Calculations in austenite are, however,
limited primarily to solute dissolution, diffusion and their influence
on the electronic structure and local
geometry\cite{JiangCarter,Gavriljuk05,Gavriljuk10,Boukhvalov07}
although calculations of vacancy-C binding have been
performed\cite{Slane04}.

In this work we present a detailed study of the energetics, kinetics
and interactions of He, C and N solutes in model austenite and
austenitic systems using DFT. A full treatment of paramagnetic
austenite and FeCrNi austenitic alloys would naturally take into
account the magnetic and composition dependence of the variables under
study and while ab initio techniques are now becoming available to
model the paramagnetic state\cite{Kormann12} and calculations in
concentrated alloys are certainly achievable\cite{Klaver07}, their
complexity precludes a broad study of all the necessary variables
relevant for radiation damage modelling. Previous studies have,
instead, either taken ferromagnetic (fm) fcc nickel (Ni) as a model
austenitic system\cite{Perfect,DomainPerfect,Tucker10} or modelled
austenite using a small set of stable, magnetically ordered states, as
in our previous work\cite{KlaverFeNiCr}. The advantage is that a more
detailed study is possible, but the level of approximation involved is
certainly not ideal and careful use should be made of the results
obtained. Here, we follow the same approach used in our previous
work\cite{KlaverFeNiCr}, performing our calculations in the two most
stable ordered magnetic states of fcc Fe. In addition, we present and
compare the results of corresponding calculations in fm Ni in order to
make more general conclusions in Fe-Ni based austenitic alloys.

In section \ref{compDetails} we present the details of our
calculations. We then proceed to present and discuss our results for
He, C and N solutes in defect free austenite and dilute Fe-Cr-Ni
austenitic alloys in section \ref{solutesInIron} and their
interactions with point defects and small vacancy clusters in section
\ref{soluteDefectSection} before making our conclusions.

\section{Computational Details}
\label{compDetails}

The calculations presented in this paper have been performed using the
plane wave DFT code, VASP\cite{KresseHafner,KresseFurthmuller}, in the
generalised gradient approximation with exchange and correlation
described by the parametrisation of Perdew and Wang\cite{PW91} and
spin interpolation of the correlation potential provided by the
improved Vosko-Wilk-Nusair scheme\cite{vwn}. Standard projector
augmented wave potentials\cite{Blochl,KresseJoubert} supplied with
VASP were used for Fe, He, C, N, Ni and Cr with 8, 2, 4, 5, 10 and 6
valence electrons respectively. First order ($N=1$) Methfessel and
Paxton smearing\cite{MethfesselPaxton} of the Fermi surface was used
throughout with the smearing width, $\sigma$, set to 0.2 eV to ensure
that the error in the extrapolated energy of the system was less than
1 meV per atom. A $2^3$ k-point Monkhorst-Pack grid was used to sample
the Brillouin zone and a plane wave cutoff of 450 eV.

All calculations used supercells of 256 ($\pm 1$, $\pm 2$,...) atoms,
with supercell dimensions held fixed at their equilibrium values and
ionic positions free to relax. For the relaxation of single
configurations, structures were deemed relaxed once the forces on all
atoms had fallen below 0.01 eV/$\angs$. For the nudged elastic
band\cite{NEB98} (NEB) calculations used to determine migration
barriers an energy tolerance of 1 meV or better was used to control
convergence. Spin-polarised calculations have been performed
throughout this work with local magnetic moments on atoms initialised
to impose the magnetic state ordering but free to relax during the
calculation. The relaxed local magnetic moments were determined by
integrating the spin density within spheres centred on the
atoms. Sphere radii of 1.302, 0.635, 0.863, 0.741, 1.286 and 1.323
$\angs$ were used for Fe, He, C, N, Ni and Cr respectively.

We have performed our calculations in both the face centred tetragonal
(fct) antiferromagnetic single layer (afmI) and double layer (afmD)
collinear magnetic reference states for austenitic Fe (at T=0 K),
which we will refer to as afmD Fe and afmI Fe in what follows,
respectively, using the same methodology as our previous
work\cite{KlaverFeNiCr}. Both of these structures consist of
(ferro-)magnetic (001) fcc planes, which we will refer to as magnetic
planes in what follows, but with opposite magnetic moments on adjacent
planes in the afmI state and an up,up,down,down ordering of moments in
adjacent magnetic planes in the afmD state. The fcc fm and fct fm
states were found to be structurally unstable and spontaneously
transformed upon addition of a whole range of defects and
solutes\cite{KlaverFeNiCr}. The fcc ferromagnetic high-spin (fm-HS)
state was, however, found to be stable to isotropic effects and we
have performed a select few calculations in this state for comparison
with other work in the literature\cite{JiangCarter}. We have also
performed a number of calculations in fcc fm Ni, which we will refer
to, simply, as Ni in what follows, where these results were not
available in the literature. We take the lattice parameters for afmI
Fe as $a=3.423\ \angs$ and $c=3.658\ \angs$, those for afmD Fe as
$a=3.447\ \angs$ and $c=3.750\ \angs$ and take $a=3.631\ \angs$ for
fm-HS Fe. Calculations in Ni have been performed with an equilibrium
lattice parameter of $a=3.522\ \angs$. The corresponding magnitudes
for the local magnetic moments in bulk, equilibrium afmI, afmD and
fm-HS Fe were determined as 1.50, 1.99 and 2.57 $\magneton$,
respectively\cite{KlaverFeNiCr} and a local moment of 0.59 $\magneton$
was found in bulk, equilibrium Ni.  Convergence tests indicated that
local moments were determined to a few hundredths of $\magneton$.

We use elastic constants for our reference states, as determined
previously\cite{KlaverFeNiCr}, or determined here using the same
techniques. For fm-HS Fe, we find $C_{11} = 40$ GPa, $C_{12} = 240$
GPa and $C_{44} = -10$ GPa, which clearly shows instability to shear
strains and tetragonal deformations, as $C^\prime = C_{11} - C_{12} =           
-200$ GPa. It is, however, stable to isotropic deformations as the
bulk modulus, $B = 187$ GPa, is positive. For Ni, we find $C_{11} =             
272$ GPa, $C_{12} = 158$ GPa and $C_{44} = 124$ GPa, which gives
$C^\prime = 114$ GPa and $B = 196$ GPa, and shows that this material
is stable to any strain deformations.

We have determined the solution enthalpy for carbon in Fe and Ni using
diamond as a reference state. The diamond structure was determined
using the same settings as our other calculations but with sufficient
k-point sampling to ensure absolute convergence of the energy. We
found a lattice parameter of $a = 3.573\ \angs$, in good agreement
with the standard experimental value.

We define the formation energy, $\eform$, of a configuration
containing $n_\mathrm{X}$ atoms of each element, X, relative to a set
of reference states for each element using 
\be 
   \eform = E-\sum_\mathrm{X} n_\mathrm{X}E^\mathrm{ref}_\mathrm{X}, 
\ee 
where $E$ is the calculated energy of the configuration and
$E^\mathrm{ref}_\mathrm{X}$ is the reference state energy for element,
X. We take the reference energies for Fe, Ni and Cr to be the energies
per atom in the bulk metal, that is Fe in either the afmI, afmD or
fm-HS states, as appropriate, Ni in its fcc fm ground state and Cr in
its bcc antiferromagnetic (afm) ground state. Details of the Fe and Ni
reference states are given above, whereas for Cr an equilibrium
lattice parameter of 2.848 $\angs$ was found with a corresponding
local moment of magnitude 0.87 $\magneton$. For He, C and N the
reference states were taken to be the non-magnetic free atom, as
calculated in VASP.

In a similar manner, we define the formation volume at zero pressure,
$V_\mathrm{f}$, of a configuration relative to the bulk metal by
\be 
   V_\mathrm{f} = V(0) - n_\mathrm{bulk} V_\mathrm{bulk}, 
\ee 

where $V(0)$ is the volume of the configuration at zero pressure,
$n_\mathrm{bulk}$ is the number of bulk (solvent) metal atoms in the
configuration and $V_\mathrm{bulk}$ is the volume per atom in the
defect-free bulk metal, which we found to be 11.138, 10.712, 11.970
and 10.918 $\angs^3$ in afmD Fe, afmI Fe, fm-HS Fe and Ni,
respectively. For our calculations, $V(0)$ was determined by
extrapolation from our calculations at the fixed equilibrium volume
using the residual pressure on the supercell and the bulk modulus for
the defect-free metal.

We define the binding energy between a set of $n$ species, $\{A_i\}$,
where a species can be a defect, solute, clusters of defects and
solutes etc., as
\be
\ebind(A_1,...,A_n)= \sum_{i=1}^n \eform(A_i) - \eform(A_1,...,A_n)
\ee
where $\eform(A_i)$ is the formation energy of a configuration
containing the single species, $A_i$, and $\eform(A_1,...,A_n)$ is the
formation energy of a configuration containing all of the
species. With this definition an attractive interaction will
correspond to a positive binding energy. One intuitive consequence of
this definition is that the binding energy of a species, $B$, to an
already existing cluster (or complex) of species, $\{A_1,\dots,A_n\}$,
which we collectively call $C$, is given by the simple formula,
\be
\ebind(B,C) = \ebind(B,A_1,\dots,A_n) - \ebind(A_1,\dots,A_n).
\ee
This result will be particularly useful when we consider the
additional binding of a vacancy or solute to an already existing
vacancy-solute complex.

We have quantified a number of uncertainties in the formation and
binding energies presented in this work. Test calculations were
performed to determine the combined convergence error from our choice
of k-point sampling and plane-wave cutoff energy. For interstitial C
and N solutes in a defect-free lattice, formation energies were
converged to less than 0.05 eV and formation energy differences, such
as migration energies, to less than 0.03 eV. For interstitial He the
convergence errors were half of those for C and N. For configurations
containing vacancies or self-interstitial defects, formation energies
were converged to 0.03 or 0.07 eV, respectively, while binding
energies were converged to 0.01 eV, except for the binding of He to a
vacancy, where the error was 0.03 eV.

The zero-point energy (ZPE) contributions to the formation energy,
which can be significant for light solute atoms, have not been
calculated in this work. We performed calculations of the ZPE for He,
C and N solutes in a number of test sites in afmD and afmI Fe, keeping
the much heavier Fe atoms fixed, which is equivalent to assuming they
have infinite mass\cite{JiangCarter}. The results showed that the ZPE
contributions were consistently around 0.10 eV in all cases, which we
take as an estimate of the ZPE error on the formation energies of
configurations containing C, N and He. The variation with site was,
however, surprisingly low at 0.01 eV, which we take as an estimate of
the ZPE error in formation energy differences, binding energies and in
the solution enthalpy for C, given that ZPE contribution in graphite
is very similar to in Fe\cite{JiangCarter}.

Performing calculations in a fixed supercell of volume, $V$, results
in a residual pressure, $P$, for which an Eshelby-type elastic
correction to the total and, therefore, formation
energy\cite{AcklandA,HanA} of $E^\mathrm{corr.} = -P^2V/2B$, can be
applied. As such, $E^\mathrm{corr.}$ also serves to indicate the
likely finite-volume error. For many of the configurations considered
here these corrections are negligible compared to other sources of
error. Where they are significant, however, their relevance is
discussed at the appropriate points in the text.

\section{Solutes in the defect free lattice}
\label{solutesInIron}

\subsection{Single solutes}

\begin{figure}
\includegraphics[width=0.8\columnwidth]{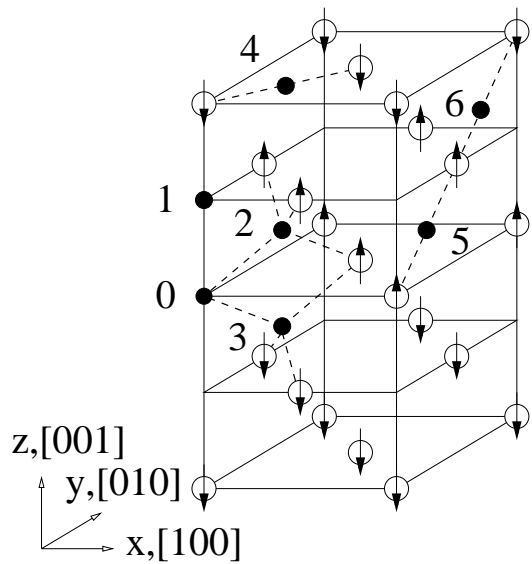}
\caption{\label{soluteFig}Substitutional (0) and interstitial
  octahedral (1), tetrahedral (2-3) and crowdion (4-6) positions (in
  black) in afmD Fe. The Fe atoms are shown in white with arrows to
  indicate the local moments. Magnetic planes are included to aid
  visualisation. The afmD Fe state, which has the lowest symmetry, is
  shown to uniquely identify all distinct positions. In afmI Fe and
  Ni, positions 2 and 3 are equivalent by symmetry, as are 5 and 6. In
  Ni, position 4 is also equivalent to 5 and 6.}
\end{figure}

The formation energies for substitutionally and interstitially sited
He, C and N solutes in the sites shown in \reffig{soluteFig} are given
in \reftab{solutesInIronTab} for Fe and \reftab{solutesInNiTab} for
Ni. We found that the Eshelby corrections were negligible for
substitutional He but could be as high as 0.02 eV in magnitude for
interstitial He and 0.04 eV for interstitial C and N. The
corresponding uncertainties in formation energy differences were
around half of these values. We discuss the results for He first,
followed by those for C and N.

\begin{table*}[thbp]
\begin{ruledtabular}
\begin{tabular}{ccccccc}
\multirow{4}{*}{Config.} & \multicolumn{2}{c}{He} & \multicolumn{2}{c}{C} & \multicolumn{2}{c}{N} \\
 & afmD Fe & afmI Fe & afmD Fe &afmI Fe & afmD Fe & afmI Fe \\
 & $\eform$ & $\eform$ & $\eform$ & $\eform$ & $\eform$ & $\eform$ \\
& ($\Delta \eform$) & ($\Delta \eform$) & ($\Delta \eform$) & ($\Delta \eform$) & ($\Delta \eform$) & ($\Delta \eform$) \\
\hline
\multirow{2}{*}{sub (0)} & {\bf 4.024} & {\bf 4.185} & -6.981 & -6.244 & \multirow{2}{*}{rlx (other)} & -5.153 \\
& (---) & (---) & (---) & (---) & & (---) \\[3pt]
\multirow{2}{*}{octa (1)} & 4.669 & 5.026 & {\bf -8.797} & {\bf -8.856} & {\bf -8.602} & {\bf -8.621} \\
 & (0.206) & (0.059) & (0.000) & (0.000) & (0.000) & (0.000) \\[3pt]
\multirow{2}{*}{tetra uu (2)} & 4.529 & as & -6.535 & as & -6.917 & as\\
 & (0.066) & tetra ud & (2.261) & tetra ud & (1.685) & tetra ud \\[3pt]
\multirow{2}{*}{tetra ud (3)} & {\bf 4.464} & {\bf 4.967} & -6.644 & -6.272 & -7.044 & -6.737 \\
 & (0.000) & (0.000) & (2.153) & (2.585) & (1.558) & (1.884) \\[3pt]
\multirow{2}{*}{$[110]$ crow. (4)} & \multirow{2}{*}{rlx (3)} & 5.271 & -6.764 & -6.412 & \multirow{2}{*}{rlx (3)} & -6.006 \\
 & & (0.303) & (2.033) & (2.445) & & (2.614) \\[3pt]
\multirow{2}{*}{$[011]$ crow. uu (5)} & 4.827 & as & -7.354 & as & -7.000 & as \\
 & (0.364) & $[011]$ crow. ud & (1.443) & $[011]$ crow. ud & (1.602) & $[011]$ crow. ud \\[3pt]
\multirow{2}{*}{$[011]$ crow. ud (6)} & 4.802 & 5.188 & -7.487 & -6.744 & -7.218 & -6.328 \\
 & (0.338) & (0.221) & (1.310) & (2.113) & (1.384) & (2.293) \\
\end{tabular}
\end{ruledtabular}
\caption{\label{solutesInIronTab} Formation energies, $\eform$, in eV,
  for substitutionally and interstitially sited solute atoms, as shown
  in \reffig{soluteFig}. The formation energies in bold are for the
  most stable states. For He, which is most stable substitutionally,
  the most stable interstitial site is also highlighted. The formation
  energy differences, $\Delta \eform$ (in brackets), to the most
  stable interstitial configurations are also given, in eV. Where the
  configuration was found to be unstable the configuration to which it
  relaxed is given. The substitutional N configuration in the fct afmD
  state relaxed to one with an octa N at 1nn to a vacancy.}
\end{table*}

\begin{table}[htbp]
\begin{ruledtabular}
\begin{tabular}{lccc}
\multirow{2}{*}{Config.} & He & C & N \\
 & $\eform$ & $\eform$ & $\eform$ \\
& ($\Delta \eform$) & ($\Delta \eform$) & ($\Delta \eform$) \\
\hline
\multirow{2}{*}{sub (0)} & {\bf 3.185} & -5.386 & -4.562 \\
& (---) & (---) & (---) \\[3pt]
\multirow{2}{*}{octa (1)} & 4.589 & {\bf -8.422} & {\bf -7.520} \\
& (0.129) & (0.000) & (0.000) \\[3pt]
\multirow{2}{*}{tetra (2-3)} & {\bf 4.460} & -6.764 & -6.497 \\
& (0.000) & (1.659) & (1.023) \\[3pt]
\multirow{2}{*}{$\langle 110\rangle$ crow. (4-6)} & 4.651 & -6.795 & -5.970 \\
& (0.191) & (1.628) & (1.550) \\[3pt]
\end{tabular}
\end{ruledtabular}
\caption{\label{solutesInNiTab} Formation energies, $\eform$, in eV,
  for substitutionally and interstitially sited He, C and N atoms in
  Ni. The layout and data content is as in \reftab{solutesInIronTab}}
\end{table}

\subsubsection{He solute}
\label{HeSoluteSection}

We found that He exhibits a large, positive formation energy in all
sites but is most stable substitutionally, which is consistent with
existing DFT studies of He in other bcc and fcc
metals\cite{Perfect,Zu,DomainPerfect,Yang,Zeng09,Fu0507,SeletskaiaA,Becquart12,Becquart06,Becquart09,Becquart10,Zhang11}. The
standard explanation is that, as a closed shell noble gas element,
bonding interactions should be primarily repulsive, leading to
insolubility and a preference for sites with the largest free
volume\cite{SeletskaiaA,Zu}. This result distinguishes He from other
small solutes, such as C and N, which are more stable interstitially
but also distinguishes it from substitutional alloying elements, such
as Ni and Cr with formation energy differences between substitutional
and interstitial sites in Fe of 3.0 eV and
above\cite{KlaverFeNiCr,OlssonA}.

In Fe, the influence of substitutional He on the local magnetic
moments of atoms in its first nearest neighbour (1nn) shell was found
to be similar to those for a vacancy, being generally enhanced
relative to the bulk moment and by up to 0.38 $\magneton$ here. This
is similar to He in bcc Fe\cite{Fu0507,SeletskaiaA}. Indeed, we found
that if the He atom was removed from the relaxed substitutional
configuration with no further relaxation, the local 1nn Fe moments
changed by less than 0.03 $\magneton$. In contrast to the vacancy,
however, where 1nn Fe were displaced inwards by 0.09 and 0.02 $\angs$
in afmD and afmI Fe, respectively\cite{KlaverFeNiCr}, the respective
displacements around a substitutional He were, on average, outwards by
0.02 and 0.04 $\angs$. This contrast can also be seen in the formation
volumes, which were found to be 0.74 $V_\mathrm{bulk}$ and 0.96
$V_\mathrm{bulk}$ for a vacancy, compared to 1.17 $V_\mathrm{bulk}$
and 1.38 $V_\mathrm{bulk}$ for substitutional He in afmD and afmI Fe,
respectively. Results in Ni were very similar to Fe, with enhanced
moments in the 1nn shell around both a vacancy and substitutional He,
a contraction of 0.04 $\angs$ in the 1nn shell around a vacancy and an
expansion of 0.02 $\angs$ around substitutional He. The formation
volume for substitutional He, at 1.02 $V_\mathrm{bulk}$, was again
found to be greater than that for the vacancy, at 0.66
$V_\mathrm{bulk}$.

The large formation energy difference, of around 2 eV, between
substitutional He and the underlying vacancy in Fe and Ni [see Section
  \ref{soluteDefectSection}], which must be due to chemical
interactions, may seem at odds with the relatively inert behaviour of
He mentioned above. However, similar results in bcc Fe have been
reproduced using simple pair potentials\cite{JuslinPot,ChenPot}, which
demonstrates that such a large energy difference, once distributed over
1nn and 2nn bonds, is commensurate with the relatively small forces
observed on the neighbouring Fe atoms around substitutional He.

In Fe, interstitial He was found to be most stable in the tetrahedral
(tetra) site, the octahedral (octa) site being the next most stable
and lying 0.206 and 0.059 eV higher in energy in the afmD and afmI
states, respectively. There is no consistent ordering of the octa and
tetra sites in ab initio studies of other fcc metals, with the octa
site being most stable in Ag\cite{Zu}, Al\cite{Yang} and
Pd\cite{Zu,Zeng09} and the tetra site being most stable in Cu\cite{Zu}
and Ni\cite{Perfect,DomainPerfect,Zu}, as our results for Ni
confirm. In both Fe and Ni, however, He favours the tetra site, which
gives a strong indication that the tetra site will also be the most
stable interstitial site in concentrated Fe-Ni based austenitic
alloys.

The other interstitial sites considered here lie no more than 0.364 eV
above the tetra site, suggesting many low energy migration paths for
interstitial He, that is, in the absence of any lattice defects that
can act as strong traps. The bilayer structure in afmD Fe breaks the
symmetry of the octa site and a He atom placed there was found to
spontaneously relax in the $[00\bar{1}]$ direction (as defined in
\reffig{soluteFig}), to between layers of the same spin by
0.55~$\angs$. It is, perhaps, surprising that in both afmI Fe and Ni,
an octa-sited He was also found to be unstable to small displacements
in many directions. We present the results of these calculations in
\reftab{octaTab}.

\begin{table}[htbp]
\begin{ruledtabular}
\begin{tabular}{ccccccc}
\multirow{3}{*}{Config.} & \multicolumn{2}{c}{afmD Fe} & \multicolumn{2}{c}{afmI Fe} & \multicolumn{2}{c}{Ni} \\
 & $\eform$ & \multirow{2}{*}{$\Delta r$} & $\eform$ & \multirow{2}{*}{$\Delta r$} & $\eform$ & \multirow{2}{*}{$\Delta r$} \\
& ($\Delta \eform$) & & ($\Delta \eform$) & & ($\Delta \eform$) & \\
\hline
\multirow{2}{*}{octa sym.} & \multicolumn{2}{c}{rlx} & 5.208 & \multirow{2}{*}{0.00} & 4.617 & \multirow{2}{*}{0.00} \\
 & \multicolumn{2}{c}{octa $[00\bar{1}]$} & (0.241) & & (0.157) & \\[3pt]
\multirow{2}{*}{octa [100]} & \multicolumn{2}{c}{rlx} & 5.105 & \multirow{2}{*}{0.39} & 4.607 & \multirow{2}{*}{0.29} \\
 & \multicolumn{2}{c}{octa $[00\bar{1}]$} & (0.138) & & (0.147) & \\[3pt]
\multirow{2}{*}{octa [001]} & 4.812 & \multirow{2}{*}{0.30} & 5.026 & \multirow{2}{*}{0.50} & \multicolumn{2}{c}{as} \\
 & (0.348) & & (0.059) & & \multicolumn{2}{c}{octa [100]} \\[3pt]
\multirow{2}{*}{octa $[00\bar{1}]$} & 4.669 & \multirow{2}{*}{0.58} & \multicolumn{2}{c}{as} & \multicolumn{2}{c}{as} \\
 & (0.206) & & \multicolumn{2}{c}{octa [001]} & \multicolumn{2}{c}{octa [100]} \\[3pt]
\multirow{2}{*}{octa [110]} & \multicolumn{2}{c}{rlx} & 5.079 & \multirow{2}{*}{0.54} & 4.589 & \multirow{2}{*}{0.54} \\
 & \multicolumn{2}{c}{octa $[00\bar{1}]$} & (0.112) & & (0.129) & \\[3pt]
\multirow{2}{*}{octa [011]} & 4.799 & \multirow{2}{*}{0.58} & 5.035 & \multirow{2}{*}{0.23} & \multicolumn{2}{c}{as} \\
 & (0.335) & & (0.068) & & \multicolumn{2}{c}{octa [110]} \\[3pt]
\multirow{2}{*}{octa $[01\bar{1}]$} & \multicolumn{2}{c}{rlx} & \multicolumn{2}{c}{as} & \multicolumn{2}{c}{as} \\
 & \multicolumn{2}{c}{octa $[00\bar{1}]$} & \multicolumn{2}{c}{octa [011]} & \multicolumn{2}{c}{octa [110]} \\[3pt]
\multirow{2}{*}{octa [111]} & \multicolumn{2}{c}{rlx} & 5.029 & \multirow{2}{*}{0.62} & \multicolumn{2}{c}{rlx} \\
 & \multicolumn{2}{c}{tetra ud} & (0.062) & & \multicolumn{2}{c}{tetra} \\[3pt]
\multirow{2}{*}{octa $[11\bar{1}]$} & \multicolumn{2}{c}{rlx} & \multicolumn{2}{c}{as} & \multicolumn{2}{c}{as} \\
 & \multicolumn{2}{c}{octa $[00\bar{1}]$} & \multicolumn{2}{c}{octa [111]} & \multicolumn{2}{c}{octa [111]} \\[3pt]
\end{tabular}
\end{ruledtabular}
\caption{\label{octaTab} Formation energies, $\eform$, and formation
  energy differences, $\Delta \eform$, in eV, to the most stable tetra
  site (in brackets) for octa-sited He atoms in Fe. He is either sited
  symmetrically (sym.)  or has been displaced off-centre, in which
  case the direction of the displacement is used to label the
  configuration and the displacement length after relaxation, $\Delta
  r$ is given, in $\angs$. The symmetrical position is as shown in
  \reffig{soluteFig} and directions determined from that point with
  the coordinate system shown. When no stable local energy minimum was
  found the state to which the configuration relaxed is given.}
\end{table}

It is particularly clear in the afmI Fe data that lower energy
configurations were found along all of our test directions, with He
relaxing to between 0.23 and 0.62 $\angs$ from the symmetrical
position. The picture is less clear in afmD Fe, where He was generally
found to relax to the lowest local energy minimum but other metastable
positions were found. In Ni, the drop in energy is far less pronounced
than in Fe but is still present, with He relaxing to stable positions
0.29 $\angs$ from the centre along $\langle 100\rangle$ directions and
0.54 $\angs$ along $\langle 110\rangle$ directions. These
configurations are important, certainly as intermediate states for the
migration of interstitial He, but also as potential transition states
and already suggest a low migration energy barrier. We study these
possibilities in detail in Section \ref{SoluteMigration}. For
completeness, we also tested for the presence of stable off-centre
positions for tetra-sited He but relaxation always returned He to the
symmetrical position.

The displacements of 1nn Fe atoms around interstitially-sited He were,
unsurprisingly, found to be greater than for the substitutional
site. A tetra-sited He in afmI Fe displaced its neighbours by 0.23
$\angs$. In afmD Fe, displacements of 0.22 and 0.32 $\angs$ were found
for tetra uu and tetra ud -sited He, respectively. The magnetic
moments on the 1nn Fe atoms were quenched relative to the bulk moments
by 0.24 $\magneton$ in afmI Fe and by 0.16 $\magneton$ for the tetra
uu site in afmD Fe but enhanced by 0.15 $\magneton$ for the tetra ud
site. We attribute this difference to the greater free volume into
which 1nn Fe atoms around a tetra ud site may be displaced. We found
formation volumes of 0.82 $V_\mathrm{bulk}$ and 0.99 $V_\mathrm{bulk}$
for tetra-sited He in afmI Fe and tetra-ud-sited He in afmD Fe,
respectively. Once again, results in Ni were similar to Fe, with a
0.24 $\angs$ displacement and moment quench of 0.09 $\magneton$ in 1nn
Ni atoms around a tetra-sited He and a formation volume of 0.78
$V_\mathrm{bulk}$.

In the most stable octa configuration in Fe, the local geometry is
complicated by the displacement of He from the symmetrical
position. For that reason, we define a local unit cell surrounding the
octa site using the positions of its six 1nn metal atoms, which lie at
the centres of the cell faces, and report on the lattice parameters of
that cell. In both afmI and afmD Fe, the local lattice parameter along
[100] and [010] directions, $a_\mathrm{1nn}$, is increased by 0.31
$\angs$ relative to the bulk equilibrium lattice, with the local
lattice parameter along the [001] direction, $c_\mathrm{1nn}$,
exhibiting an increase of 0.26 $\angs$ in the afmD state and 0.29
$\angs$ in the afmI state. The local moment on the 1nn Fe atom that He
is displaced towards is significantly quenched by 1.04 and 0.41
$\magneton$ in the afmD and afmI states, respectively. In contrast,
the other 1nn moments are moderately enhanced by between 0.03 and 0.17
$\magneton$. In Ni, the most stable off-centre octa position is along
$\langle 110\rangle$ directions from the symmetrical position. The
resulting local unit cell, which exhibits a very slight shear, has
$c_\mathrm{1nn} \ne a_\mathrm{1nn}$, with $a_\mathrm{1nn}$ increased
by 0.31 $\angs$ relative to bulk and $c_\mathrm{1nn}$ by 0.28
$\angs$. Local 1nn Ni moments were found to be quenched by between
0.02 and 0.08 $\magneton$.

These findings suggest that the relative stability of tetra over octa
He, which is opposite to the order suggested by free volume
arguments\cite{SeletskaiaA,Zu}, may be best ascribed to the relative
ease with which a tetra He may lower its purely repulsive interactions
with neighbouring atoms by local dilatation. To further investigate
this hypothesis in Fe we split the formation energy for unrelaxed and
relaxed substitutional, octa and tetra He configurations into three
terms, in a similar manner to the work of Fu {\it et
  al.}\cite{Fu08}. The first is the formation energy,
$\eform^\mathrm{def.}$, of any underlying, atomically relaxed,
defects, e.g. a single vacancy for substitutional He. The second is
the mechanical energy, $\eform^\mathrm{mech.}$, required to deform the
Fe matrix containing those relaxed defects to the exact positions
found in the configuration under study. The third is the energy change
from chemical interactions, $\eform^\mathrm{chem.}$, upon insertion of
the solute into its final position with no further relaxation. We also
define the insertion energy, $\eform^\mathrm{ins.}$, as the sum of
$\eform^\mathrm{mech.}$ and $\eform^\mathrm{chem.}$ i.e. the formation
energy for insertion of a solute into any position in a relaxed Fe
matrix containing any relevant defects. We take the insertion energy
as a more appropriate measure of site preference than the (total)
formation energy, $\eform$. The results are given in
\reftab{EformTab}.

\begin{table}[htbp]
\begin{ruledtabular}
\begin{tabular}{cccc}
 Config. & $\eform^\mathrm{mech.}$ & $\eform^\mathrm{chem.}$ & $\eform^\mathrm{ins.}$ \\
\hline
 \multicolumn{4}{c}{afmD Fe + He} \\
\hline
sub, unrelaxed & 0.136 & 2.150 & 2.286 \\
sub, relaxed & 0.167 & 2.045 & 2.212 \\
tetra, unrelaxed & 0.000 & 6.778 & 6.778 \\
tetra, relaxed & 1.330 & 3.134 & 4.464 \\
octa, unrelaxed & 0.000 & 5.804 & 5.804 \\
octa, relaxed & 0.755 & 3.914 & 4.669 \\
\hline
 \multicolumn{4}{c}{afmI Fe + He} \\
\hline
sub, unrelaxed & 0.023 & 2.662 & 2.685 \\
sub, relaxed & 0.155 & 2.073 & 2.228 \\
tetra, unrelaxed & 0.000 & 6.774 & 6.774 \\
tetra, relaxed & 0.999 & 3.968 & 4.967 \\
octa, unrelaxed & 0.000 & 6.081 & 6.081 \\
octa, relaxed & 0.855 & 4.171 & 5.026 \\
\end{tabular}
\end{ruledtabular}
\caption{\label{EformTab} Mechanical deformation energy,
  $\eform^\mathrm{mech.}$, and chemical bonding energy,
  $\eform^\mathrm{chem.}$, contributions to the total formation
  energy, $\eform$, and the insertion energy, $\eform^\mathrm{ins.}$,
  for unrelaxed and relaxed substitutional, tetra and octa He solute
  configurations in afmD and afmI Fe, in eV. The most stable octa
  configuration was used in both states and the tetra ud configuration
  was used for the afmD state.}
\end{table}

The substitutional site is clearly the most favoured, even in the
unrelaxed state and by at least 2.25 eV once relaxed. In the unrelaxed
lattice, an octa He is significantly more stable than a tetra He, as
expected from purely repulsive interactions given the relative
proximity of 1nn Fe in the two sites. Under relaxation the chemical
bonding energy is significantly reduced and to a far greater degree in
the tetra site. The positive mechanical deformation energy is also
greater for tetra He but the net result is still to stabilise tetra
over octa He. These results clearly show that the relative stability
of He in tetra and octa sites can be understood as resulting from a
balance between the energy required for local dilatation of the Fe
matrix coupled with a purely repulsive Fe-He interaction, which we
suggest could be easily modelled using a simple pair potential.

In bcc Fe, the relative stability of tetra over octa He has been
explained as resulting from strong hybridisation of He p-states with
Fe d-states\cite{SeletskaiaA,Zu}. However, we do not find the evidence
for such strong hybridisation to be convincing. We suggest that a
repulsive non-bonding mechanism also applies to bcc Fe and explains
the difference in a much simpler manner. The magnetic and polarisation
effects discussed by Seletskaia\cite{SeletskaiaA} and Zu\cite{Zu} are
a simple consequence of these non-bonding interactions and not He
p-state, Fe d-state hybridisation. Formation energy
calculations\cite{SeletskaiaA,Zu} show that octa-sited He is higher in
energy both before and after relaxation, despite the relaxation energy
for octa He being greater than for tetra He. This results, primarily,
from the very short 1nn Fe-He separations in the octa site when
compared to those for the tetra site and the relative strengths of the
resulting repulsive interactions. The fact that purely repulsive pair
potentials for Fe-He interactions in bcc Fe are capable of reproducing
the relative stability\cite{JuslinPot,ChenPot} gives further support
to our claim.

\subsubsection{C and N solutes}

The results for C and N solutes [in
  \reftabs{solutesInIronTab}{solutesInNiTab}] show that both elements
clearly favour the octa interstitial site in both Fe and
Ni. Experimental observations show this to be the preferred site for C
in an Fe-13wt\%Ni-1wt\%C austenitic alloy\cite{Butler}. One exception
worth comment is that of substitutional C in afmD Fe, for which the
insertion energy, $\eform^\mathrm{ins.}$, which as discussed for He
provides a more appropriate measure of site preference, is comparable
to that for octa C. On further inspection we found that, due to the
asymmetries in the afmD state, the initially on-lattice C atom relaxed
to 0.77 $\angs$ from the lattice site. While this displacement is
certainly significant, the C atom remains closer to the substitutional
site than to an octa position at 1nn to the (vacated) lattice site and
has been named to reflect this difference. Relaxation of the
substitutional N configuration also resulted in displacement away from
the lattice site but convergence was to a configuration with the N
atom in an octa site at 1nn to a vacancy. We performed calculations to
test for the presence of any stable off-centre octa configurations for
C and N but none were found, in contrast to the results for He.

\begin{table*}[thbp]
\begin{ruledtabular}
\begin{tabular}{ccccccccc}
& \multicolumn{2}{c}{afmD Fe} & \multicolumn{2}{c}{afmI Fe} & \multicolumn{2}{c}{fm-HS Fe} & \multicolumn{2}{c}{Ni}\\ 
& C & N & C & N & C & N & C & N \\ 
\hline 
$\Delta a_\mathrm{1nn}$ & 0.321 & 0.303 & 0.305 & 0.276 & 0.174 & 0.145 & 0.183 & 0.170 \\
$\Delta c_\mathrm{1nn}$ & 0.080 & 0.048 & 0.154 & 0.127 & & & & \\ 
$c_\mathrm{1nn}/a_\mathrm{1nn}$ & 1.016 & 1.013 & 1.023 & 1.023 & & & & \\
$\Delta a/(a x_\mathrm{X}^\mathrm{f})$ & 0.266 & 0.265 & 0.341 & 0.327 & 0.072 & 0.034 & 0.243 & 0.263 \\ 
$\Delta c/(c x_\mathrm{X}^\mathrm{f})$ & -0.057 & -0.044 & 0.026 & 0.042 & & & & \\ 
$\Delta a_\mathrm{eff.}/(a_\mathrm{eff.} x_\mathrm{X}^\mathrm{f})$ & 0.158 & 0.162 & 0.236 & 0.232 & & & & \\ 
$V_\mathrm{f}/V_\mathrm{bulk}$ & 0.53 & 0.54 & 0.78 & 0.76 & 0.214 & 0.102 & 0.73 & 0.79 \\ 
$E_\mathrm{f,G}^\mathrm{sol.}$ & 0.323 & & 0.263 & & -0.164 & & 0.697 & \\
\end{tabular}
\end{ruledtabular}
\caption{\label{octaCandNTab} Lattice parameter differences ($\Delta
  a_\mathrm{1nn}$ and $\Delta c_\mathrm{1nn}$, in $\angs$) between
  those for the unit cell surrounding octa C and N solutes
  ($a_\mathrm{1nn}$ and $c_\mathrm{1nn}$) and the bulk equilibrium
  lattice parameters and the local $c_\mathrm{1nn}/a_\mathrm{1nn}$
  ratio. Linear expansion coefficients ($\Delta a/(a
  x_\mathrm{X}^\mathrm{f})$ and $\Delta c/(c
  x_\mathrm{X}^\mathrm{f})$) for the dependence of the lattice
  parameters on the fractional atomic solute composition,
  $x_\mathrm{X}^\mathrm{f}$, for solute X. For afmD and afmI Fe, the
  linear expansion coefficient for an effective lattice parameter,
  defined by $a_\mathrm{eff.} = (a^2c)^{1/3}$ is also
  given. Fractional formation volumes, $V_\mathrm{f}/V_\mathrm{bulk}$
  for octa-sited C and N solutes are given. The solution energy,
  $E_\mathrm{f,G}^\mathrm{sol.}$, taken to dissolve graphite in each
  of the reference states is given, in eV. For comparison, our
  calculations in bcc fm Fe give $E_\mathrm{f,G}^\mathrm{sol.} =
  0.700$ eV. }
\end{table*}

We discuss the influence of octa C and N solutes on the local lattice
geometry in an identical manner to octa-sited He, that is, using
$a_\mathrm{1nn}$ and $c_\mathrm{1nn}$. The results are presented in
\reftab{octaCandNTab} for both Fe and Ni, including results in fm-HS
Fe, which was shown to be mechanically unstable in our previous
work\cite{KlaverFeNiCr}, but not to the isotropic strain exerted
locally by an octa-sited solute. We include this extra state here to
compare with the work of Jiang and Carter\cite{JiangCarter}. It is
immediately clear that the geometrical influence of octa C is rather
similar to octa N, although with slightly smaller dilatations for
N. Local expansion is observed in all our reference states, although
the expansion of $c$ in afmD and afmI Fe is much less than for $a$. As
a result, the local $c/a$ ratio is significantly reduced relative to
the bulk material, to 1.02 around a C solute in both afmD and afmI Fe,
which is in good agreement with the 3\% tetragonal distortion found by
Boukhvalov {\it et al.}\cite{Boukhvalov07}, and to 1.01 and 1.02
around an N solute in afmD and afmI Fe, respectively.

The magnetic influence of octa C and N solutes is, again, very similar
with significant quenching of the local moments on 1nn solvent atoms
seen in all reference states, as expected for magnetic atoms under
compression. In both afmD and afmI Fe the effect is most pronounced in
those neighbours lying within the same magnetic plane as the solute,
which also show the most significant displacement, resulting in a
quench of 0.72(0.66) $\magneton$ for C(N) in afmD Fe and of 1.25(1.37)
$\magneton$ in afmI Fe. In fm-HS Fe, 1nn moments are quenched by
0.48(0.57) $\magneton$ around C(N) and in Ni a quench of 0.42
$\magneton$ was observed for both C and N.

In addition to this local influence, we have investigated the
dependence of the lattice parameters of our reference states on the
fractional atomic compositions, $x_\mathrm{C}^\mathrm{f}$ and
$x_\mathrm{N}^\mathrm{f}$ for C and N, respectively. For low
concentrations, as studied here, the lattice parameters change
linearly as a function of the fractional composition\cite{Cheng90}. In
this case, quantities such as $\Delta a/(a x_\mathrm{C}^\mathrm{f})$,
where $\Delta a$ is the difference between the lattice parameter with
and without solute atoms present, are dimensionless constants that
completely specify the linear expansion. Our calculations have been
performed in supercells at the equilibrium lattice parameters, so we
determine the linear expansion coefficients by extrapolating to zero
stress using the residual stress that builds up on the supercell upon
addition of a solute and a knowledge of the elastic constants [see
  Section \ref{compDetails}]. In afmD and afmI Fe we have also
calculated the linear expansion coefficients for an effective lattice
parameter, $a_\mathrm{eff.} = (a^2c)^{1/3}$, as a means to compare
more directly with experiment.

The results [in \reftab{octaCandNTab}] show that local expansion
around the solutes leads to a net expansion of the cell, overall. The
afmD state of Fe does, however, exhibit a small contraction in $c$ and
the afmI state shows very little expansion in $c$, when compared to
that for $a$. Once again, this shows that the addition of C and N acts
to reduce the $c/a$ ratio, bringing the lattice back toward perfect
fcc. In austenite, experimental results by Cheng {\it et
  al.}\cite{Cheng90} and presented by Gavriljuk {\it et
  al.}\cite{Gavriljuk05} show that $\Delta a/(a
x_\mathrm{C}^\mathrm{f})$ lies between 0.199 and 0.210, with $\Delta
a/(a x_\mathrm{N}^\mathrm{f})$ being slightly greater at between 0.218
and 0.224. Our results in afmD and afmI Fe are in broad agreement with
these values but do not differentiate between C and N. Results in
fm-HS Fe are significantly different from experiment, which again
shows the unsuitability of this state for modelling austenite. It is
interesting to note that our results for Ni are consistent with those
for austenite and do exhibit more expansion due to N than for C.
Experimental results in austenitic FeCrNi alloys\cite{Gavriljuk05} are
comparable to those for pure austenite and the general agreement with
our results strengthens the case for using afmD and afmI Fe or using
Ni as model systems for austenite and austenitic alloys.

We have determined the solution energy at fixed equilibrium volume,
$E_\mathrm{f,G}^\mathrm{sol.}$, taken to dissolve graphite into our
four bulk states [\reftab{octaCandNTab}]. We have done this by
calculating the solution energy relative to diamond and then applying
the commonly accepted experimental energy difference of 20 meV/atom
between the cohesive energies of diamond and graphite at $T=0$ K
\cite{Kormann11}. For comparison, we have calculated
$E_\mathrm{f,G}^\mathrm{sol.}$ for C in bcc fm Fe and find a value of
0.70 eV, which is in good agreement with the experimental value of
between 0.60 and 0.78 eV found by Shumilov {\it et
  al.}\cite{JiangCarter,Shumilov73}.

The solution energies in all three Fe states are significantly lower
than for bcc fm Fe. This is consistent with the relatively higher
solubility of C in austenite than in ferrite and with the well known
experimental result that C stabilises austenite over ferrite, as seen
in the phase diagram. The effect is most pronounced in the fm-HS
state, where the reaction is exothermic, in good agreement with
previous DFT calculations\cite{JiangCarter}. In combination with the
results discussed above, this implies that at sufficient
concentrations, C will act to stabilise the fm-HS state over the
others, just as was found for Ni in fcc Fe-Ni
alloys\cite{Abrikosov07}. The same conclusions follow for N by a
direct comparison of the formation energies for octa sited N [see
  \reftab{solutesInIronTab}], for which we found values of -8.252 and
-9.018 eV in bcc fm and fcc fm-HS Fe, respectively.

Experimental results for the solution enthalpy of C in
austenite\cite{Lobo76} yield a value of $E_\mathrm{f,G}^\mathrm{sol.}
= 0.37 eV$ at the concentration studied here, which agrees to within
0.1 eV with our calculations in afmD and afmI Fe but not with those
for the fm-HS state and again supports their suitability as reference
states for paramagnetic austenite. Our calculations in the
ferromagnetic state for Ni are in good agreement with previous DFT
calculations of Siegel and Hamilton\cite{Siegel03}. However, as they
report, this value is higher than those found experimentally in
high-temperature, paramagnetic Ni, which lie between 0.42 and 0.49
eV. It is worth noting that their calculations in non-magnetic (nm)
Ni, which they use to model the paramagnetic state, underestimate the
experimental range at between 0.2 and 0.35 eV. We conclude that the
calculated solution enthalpy for C in Ni is particularly sensitive to
the underlying magnetic state.

\subsection{Solute migration}
\label{SoluteMigration}

As a first step in the calculation of migration energies for He, C and
N solutes we investigated whether a 32 atom cell would be sufficient
for this purpose. To do this we recalculated the formation energies
for substitutional and interstitial He and C in afmD Fe using a 32
atom cell. We compare these with our 256 atom cell calculations
[\reftab{solutesInIronTab}] in \reftab{solutesInIronTab32}.

\begin{table}[htbp]
\begin{ruledtabular}
\begin{tabular}{lcccc}
\multirow{2}{*}{Config.} & \multicolumn{2}{c}{He} & \multicolumn{2}{c}{C} \\
 & 32 atom & Error & 32 atom & Error \\
 & $\eform$ & $\eform$ & $\eform$  &  $\eform$ \\

& ($\Delta \eform$) & ($\Delta \eform$) & ($\Delta \eform$) & ($\Delta \eform$) \\
\hline
\multirow{2}{*}{Sub (0)} & {\bf 4.039} & {\bf 0.015} & -6.911 & 0.070 \\
& (---) & (---) & (---) & (---) \\[3pt]
\multirow{2}{*}{octa (1)} & 4.730 & 0.061 & {\bf -8.798} & -0.001 \\
& (0.151) & (-0.055) & (0.000) & (0.000) \\[3pt]
\multirow{2}{*}{tetra uu (2)} & 4.607 & 0.078 & -6.395 & 0.140 \\
& (0.028) & (-0.038) & (2.403) & (0.142) \\[3pt]
\multirow{2}{*}{tetra ud (3)} & {\bf 4.579} & {\bf 0.115} & -6.544 & 0.100 \\
& (0.000) & (0.000) & (2.255) & (0.102) \\[3pt]
\multirow{2}{*}{$[110]$ crow. (4)} & \multirow{2}{*}{rlx (3)} & \multirow{2}{*}{rlx (3)} & -6.396 & 0.368 \\
& & & (2.402) & (0.369) \\[3pt]
\multirow{2}{*}{$[011]$ crow. uu (5)} & 4.897 & 0.070 & -7.209 & 0.145 \\
& (0.318) & (-0.046) & (1.589) & (0.146) \\[3pt]
\multirow{2}{*}{$[011]$ crow. ud (6)} & 4.866 & 0.064 & -7.346 & 0.141 \\
& (0.287) & (-0.051) & (1.452) & (0.142) \\[3pt]
\end{tabular}
\end{ruledtabular}
\caption{\label{solutesInIronTab32} Comparison between calculations in
  32 atom and 256 atom cells in afmD Fe of the formation energies,
  $\eform$, in eV, for substitutional and interstitial He and C
  solutes and formation energy differences, $\Delta\eform$, in eV, to
  the most stable interstitial configurations, highlighted in
  bold. The layout and data content of each column is as in
  \reftab{solutesInIronTab}. The column headed, 32 atom, contains the
  results for the 32 atom cell and the column headed, Error, contains
  the difference between the 32 atom and 256 atom results, which we
  take as an estimate of the finite volume error in the 32 atom cell.}
\end{table}

There is a significant size effect on the formation energies in the 32
atom cell, except for substitutional He and octa C, where the
formation energies are within errors of those in the 256 atom
cell. The formation energies are greater in the 32 atom cell, as
expected from volume-elastic effects, by between 0.06 and 0.12 eV for
interstitial He and by between 0.00 and 0.37 eV for interstitial
C. Formation energy differences to the most stable interstitial
configuration also exhibit a significant size effect, with the smaller
cell underestimating them by between 0.04 and 0.06 eV for He
configurations and overestimating them by between 0.10 and 0.37 eV for
C. It is reasonable to assume that the migration energy, which is
itself a formation energy difference, will suffer from similar size
effects.

For C, the choice of cell size actually changes the relative stability
of the [110] crowdion and tetra ud configurations. This is important
as these two are transition states on two distinct migration paths for
C (as will be shown in what follows). The small cell would, therefore,
give the wrong minimum energy path (MEP), as found previously for C in
fm-HS Fe\cite{JiangCarter}. Closer inspection of the [110] crowdion
configuration showed that the periodic boundary conditions in the
smaller cell applied unphysical constraints on the displacements of Fe
atoms at 1nn to C and along the crowdion axis generally, which
resulted in a significant buckling, moving the C atom towards the
tetra uu site, that is along [001], by 0.71 $\angs$. In the larger
cell these constraints are not present, resulting in a significantly
lower formation energy and while there is still a small displacement
towards the tetra uu site of 0.18 $\angs$ this is to be expected given
the asymmetry present in the afmD state.

As a final test, we investigated the case of C in fm-HS Fe, where
Jiang and Carter have determined a migration barrier in a 32 atom
cell\cite{JiangCarter}. They found that the $\langle 110\rangle$
crowdion site is an intermediate site for C migration, lying only 0.01
eV below the transition state energy and 0.98 eV above the stable octa
site. Our calculations in a 32 atom cell agree well with this finding,
with an energy difference of 1.01 eV between the $\langle 110\rangle$
crowdion and octa sites for C. However, when we repeated the
calculations in a 256 atom cell, we found that a configuration with C
in the $\langle 110\rangle$ crowdion was structurally unstable and
spontaneously transformed as a result of the non-isotropic stress on
the Fe lattice. By contrast, the isotropic stress from an octa-sited C
only led to local relaxation of the Fe matrix and maintenance of the
crystal structure. We conclude that the 32 atom cell effectively
imposed artificial constraints that allowed a, seemingly, sensible
migration barrier to be determined.

Overall, we find that the finite size effects in the 32 atom cell are
too significant and while some intermediate cell size between the two
investigated here may be sufficient, we have performed our migration
energy calculations in the 256 atom cell.

\subsubsection{Interstitial He migration}
\label{IntHeMig}

\begin{figure}
\includegraphics[width=0.7\columnwidth]{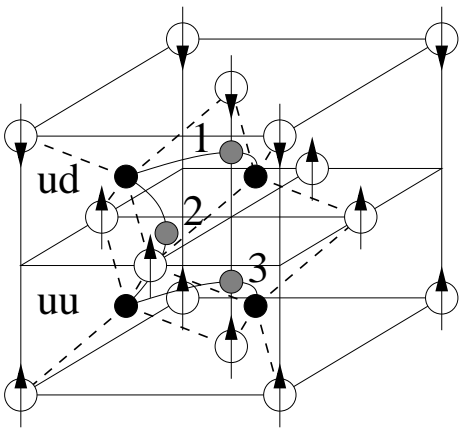}
\caption{\label{tetraHeMigFig}Possible migration paths for
  interstitial He in the afmD Fe lattice. Paths are shown for 1nn
  jumps from initial to final tetra positions (black circles) via
  off-centre octa intermediate transition state positions (grey
  circles). The Fe atoms (white circles) are shown with arrows to
  indicate the local moments. The symmetry of the afmD state leads two
  distinct tetra sites (uu and ud) and three distinct 1nn jumps, as
  shown. In the afmI state paths 1 and 3 are equivalent but still
  distinct from path 2. Coordinate axes are as in \reffig{soluteFig}.}
\end{figure}

The migration of interstitial He is relevant in the initial stages
after He production by transmutation, $\alpha$-particle irradiation
and at sufficiently high temperatures for He to escape from defect
traps. The migration of He between adjacent tetra sites (that is,
between sites at 1nn on the cubic sub-lattice of tetra sites) can
proceed along many distinct paths, with their corresponding transition
states defining the energy barrier for the transition. A direct path
would lead to an intermediate state with He in the crowdion position
but the energy differences to the tetra configurations (in
\reftabs{solutesInIronTab}{octaTab}) suggest that the direct path is
not the MEP and that the transition state has He in an off-centre
position. We show representative paths for the three distinct 1nn
jumps in afmD Fe in \reffig{tetraHeMigFig}.

\begin{figure}
\includegraphics[width=\columnwidth]{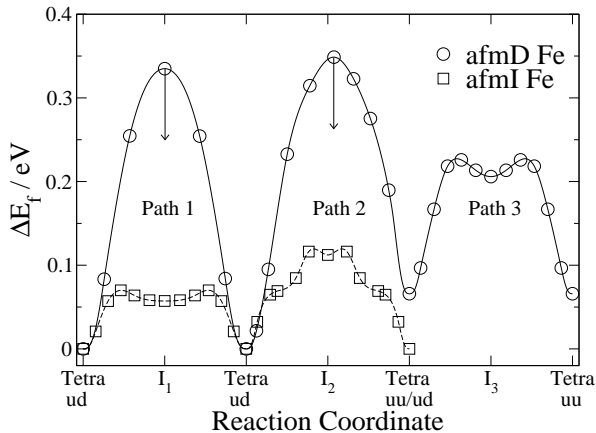}
\caption{\label{tetraHeMigGraphFig}Formation energy difference,
  $\Delta\eform$, to the lowest energy tetra configuration along the
  distinct migration paths for interstitial He in Fe, as shown in
  \reffig{tetraHeMigFig}. Positions of the tetra configurations and
  the intermediate configurations, I$_i$, along path $i$ are
  labelled. In the afmI state, the data for path 3 has been omitted as
  it is equivalent to path 1. The arrows indicate the expected
  lowering of the migration barrier heights if a reorganisation of
  magnetic moments is allowed along the migration path.}
\end{figure}

\begin{table}[htbp]
\begin{ruledtabular}
\begin{tabular}{cccc}
 Path, $i$ & afmD Fe & afmI Fe & Ni \\
\hline
1  & 0.335 & 0.070 & 0.129 \\
2 (ud to uu) & 0.349 & 0.119 & --- \\
2 (uu to ud) & 0.283 & --- & --- \\
3  & 0.160 & --- & --- \\
\end{tabular}
\end{ruledtabular}
\caption{\label{IntHeMigTab}Migration energy barrier height, $\emig$,
  in eV, for the migration of interstitial He along the distinct paths
  identified in \reffig{tetraHeMigFig}. In afmD Fe, path 2 is
  asymmetrical and the direction of migration has been identified by
  the initial and final tetra sites. In Ni, all paths are equivalent
  and the migration energy is given by the formation energy difference
  between the octa [110] and tetra He configurations from
  \reftabs{solutesInNiTab}{octaTab}.}
\end{table}

We have performed NEB calculations for He migration in Fe along these
paths and show the formation energy difference to the most stable
interstitial configuration against a, suitably chosen, reaction
coordinate in \reffig{tetraHeMigGraphFig}, with the corresponding
migration barrier heights given in \reftab{IntHeMigTab}. It is
immediately obvious that the results for the two Fe reference states
differ significantly. This is not surprising, however, given that
typical magnetic effects can be of the same order of magnitude as the
migration barrier height [see \reftab{solutesInIron}]. The high
barriers along paths 1 and 2 in afmD Fe are because the lowest-energy
tetra site is between layers of unlike moment and so not adjacent to
the lowest energy octa intermediate site, which lies off-centre between
like-spin layers [\reftab{octaTab}]. A wholesale reorganisation of
spins would lower the barriers along these paths and would be
preferred in the paramagnetic state. This problem is not present for
path 3, resulting in a significantly lower barrier, which is more
consistent with those found in the afmI state, where a more uniform
distribution of energies around the octa site exists [see
  \reftab{octaTab}]. Path 1 in the afmI state and path 3 in the afmD
state both show a double peaked structure with weakly stable octa
$[001]$ and octa $[00\bar1]$ intermediate states, respectively. These
intermediates are equidistant from four tetra sites, resulting in the
same energy barrier for 1nn and 2nn jumps on the tetra
sub-lattice. The same cannot be said for migration along path 2, which
proceeds via a (near-)octa [110] transition state in both reference
states. In the afmI state, there appears to be a very shallow minimum
at I$_2$, that is the off-centre octa [110] configuration, but with a
depth of 0.007 eV this may well be just an artefact of the convergence
criteria as it is less than the expected error for formation energy
differences. The data also exhibits a shoulder between the tetra ud
and I$_2$ sites, which we suggest results from close proximity to the
octa [111] configuration. It seems reasonable to suggest that the
barriers for 2nn jumps that cross a magnetic plane will be close in
energy to those for path 2, given the additional data in
\reftab{octaTab}. Overall, our findings suggest an energy barrier for
interstitial He migration in austenite that is below 0.35 eV and more
likely in the region between 0.1 and 0.2 eV.

Such low migration barriers are typical of all metals for which data
is available. Ab initio calculations find a value of 0.10 eV for fcc
Al\cite{Yang}, 0.07 eV for fcc Pd\cite{Zeng09} and 0.06 eV for bcc
Fe\cite{Fu0507}, W\cite{Becquart06} and V\cite{Zhang11}. Experimental
validation of these results is not forthcoming, primarily due to the
low temperatures involved and the complications of He interactions
within the material. In bcc W, Wagner and Seidman\cite{Wagner79}
estimate the migration enthalpy to be between 0.24 and 0.32 eV, with
He being immobile up to temperatures of at least 90K, which is
consistent with the value of 0.28 eV found for $^3$He migration by
Amano and Seidman\cite{Amano84}. The discrepancy between ab initio and
experiment was explained by Becquart and Domain as being due to the
presence of strong He-He binding, as found in their ab initio
calculations, resulting in the formation of less mobile interstitial
He clusters for all but the lowest
concentrations\cite{Becquart06}. This is consistent with the work of
Soltan {\it et al.}\cite{Soltan91}, who found He to be mobile in W and
Au at temperatures below 5K but with increasing suppression of
mobility as a function of He concentration.

To this data we add the results of our own investigation into He
migration in Ni. Following on from the results in afmD and afmI Fe, we
make the reasonable assumption that the most stable off-centre octa He
configuration is a good candidate for the transition state for
interstitial He migration. The additional uncertainty on the inferred
migration barrier height from this assumption should be less than 0.01
eV. From the results presented in \reftab{octaTab} this is the
off-centre octa $\langle 110\rangle$ configuration, with a
corresponding migration barrier height of 0.13 eV, which compares well
with the experimental value of $0.14\pm 0.03$ eV measured by Philipps
and Sonnenberg\cite{Philipps83}, corresponding to a migration
activation temperature of $55\pm 10$ K. This barrier height also
compares well with our best estimate for austenitic Fe.  We therefore
tentatively suggest that the barrier height for interstitial He
migration in austenitic Fe-Ni based alloys will be in the range 0.1 to
0.2 eV, resulting in free, three-dimensional diffusion well below room
temperature. We accept that there is a very real possibility of
significant local composition dependence in these concentrated alloys
but we speculate that the effective barrier height will still be in
the given range.

\subsubsection{Substitutional He migration}
\label{SubHeMigSection}

The diffusion of substitutional He generally proceeds via the
dissociative and vacancy
mechanisms\cite{Sciani83,Mansur86,Adams88}. Direct exchange with a
neighbouring solute atom provides an alternative
mechanism\cite{Adams88} but is highly unlikely to contribute
significantly to diffusion due to the large activation energy for the
process. For example, our best estimate of the barrier height in Ni is
3.50 eV, which compares well with that found using an embedded atom
model (EAM) potential of 3.1 eV by Adams and Wolfer\cite{Adams88} and
means that substitutional He is, essentially, immobile.

For many applications, substitutional He is best considered as an
interstitial He atom strongly bound to a vacancy point defect, with a
binding energy, $\ebind(\mathrm{He}^\mathrm{I},\mathrm{V})$. The
dissociative mechanism for substitutional He migration proceeds by the
dissociation of He from a vacancy followed by interstitial migration
until it becomes trapped in another vacancy. As such, the diffusion
coefficient by this mechanism is inversely proportional to the vacancy
concentration\cite{Sciani83,Adams88,Vassen91}. If thermal vacancies
dominate, the activation energy is given
by\cite{Sciani83,Mansur86,Adams88,Vassen91} 
\be
   E_\mathrm{A}^\mathrm{diss.} = \emig(\mathrm{He}^\mathrm{I}) + \ebind(\mathrm{He}^\mathrm{I},\mathrm{V}) - \eform(\mathrm{V}),
\label{EmDiss1}
\ee
where $\emig(\mathrm{He}^\mathrm{I})$ is the migration energy for
interstitial He. However, if there is a supersaturation of vacancies,
for example under irradiation, then the diffusion is dominated by the
dissociation step and
\be 
E_\mathrm{A}^\mathrm{diss.} = \emig(\mathrm{He}^\mathrm{I}) + \ebind(\mathrm{He}^\mathrm{I},\mathrm{V}),
\label{EmDiss2}
\ee 
which is, essentially, the dissociation energy for substitutional He
from its vacancy, and the diffusion coefficient will remain inversely
proportional to the vacancy concentration\cite{Adams88,Vassen91}.

The diffusion of a substitutional solute by the vacancy mechanism in
an fcc lattice is usually well described by the 5-frequency model of
Lidiard and LeClaire\cite{Lidiard55,LeClaire56}. A key assumption of
this model is that when a vacancy binds at 1nn to a substitutional
solute, the solute remains on-lattice. However, this is not the case
for He, which we find relaxes to a position mid-way between the two
lattice sites to form a V$_2$He complex. The possibility of
solute-vacancy exchange at 2nn is also not included in this model, a
point we will return to in the following discussion.

Given the strong binding between a vacancy and substitutional He at
1nn, which we discuss in Section \ref{soluteVacancySection}, we assume
that the migration of the V$_2$He complex, as a single entity,
dominates the diffusion by the vacancy mechanism\footnote{Other
  processes may contribute and a model, generalising the
  five-frequency model of Lidiard and LeClaire, is needed for this
  situation. }, with a migration energy,
$\emig(\mathrm{V}_2\mathrm{He})$. The diffusion coefficient will be
proportional to the V$_2$He concentration, which is in turn
proportional to the vacancy concentration and depends on the binding
energy between a substitutional He and a vacancy,
$\ebind(\mathrm{He}^\mathrm{S},V)$. The resultant activation energy
for substitutional He migration by the vacancy mechanism is given
by\cite{Sciani83}, \be E_\mathrm{A}^\mathrm{vac.} =
\emig(\mathrm{V}_2\mathrm{He}) - \ebind(\mathrm{He}^\mathrm{S},V) +
\eform(\mathrm{V})
\label{EmVac1}
\ee
when thermal vacancies dominate and by
\be
E_\mathrm{A}^\mathrm{vac.} = \emig(\mathrm{V}_2\mathrm{He}) - \ebind(\mathrm{He}^\mathrm{S},V)
\label{EmVac2}
\ee 
when there is a supersaturation of vacancies\cite{Fu0507}. 

\begin{figure}
\includegraphics[width=0.7\columnwidth]{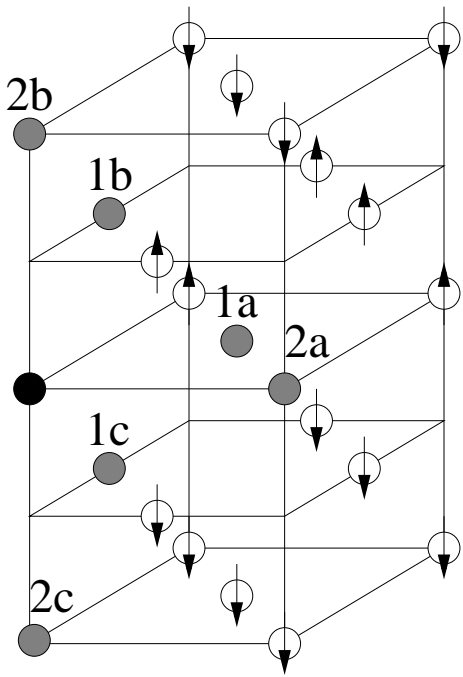}
\caption{\label{SubSubFig}Configurations for A-B pairs of interacting
  substitutionally sited solutes and defects in afmD Fe. Species A is
  shown in black and species B in grey along with the configuration
  label. Fe atoms are shown in white with arrows to indicate the local
  moments. Coordinate axes are as in \reffig{soluteFig}.}
\end{figure}

We have determined the migration energies for the V$_2$He complex
using a combination of NEB and single configuration calculations.  In
afmD and afmI Fe, where more than one distinct V$_2$He complex exists,
we have calculated the migration energy along all of the distinct
paths where the migrating Fe atom retains the sign of its magnetic
moment. In previous work\cite{KlaverFeNiCr}, we found unrealistically
high migration barriers along paths where the moment changed. We label
the migration paths for V$_2$He migration by the initial and final
configurations, which are defined in \reffig{SubSubFig}, and present
the corresponding migration energies in \reftab{HeV2MigTab}.

\begin{table}[htbp]
\begin{ruledtabular}
\begin{tabular}{cccc}
Path & afmD Fe & afmI Fe & Ni \\
\hline
1b$\rightarrow$1b & 1.033 & as 1c$\rightarrow$1c & 1.197 \\
1c$\rightarrow$1c & 0.910 & 0.898 & --- \\
1a$\rightarrow$1b & 1.216 & --- & --- \\
1b$\rightarrow$1a & 1.211 & --- & --- \\
\end{tabular}
\end{ruledtabular}
\caption{\label{HeV2MigTab}Migration energies,
  $\emig(\mathrm{V}_2\mathrm{He})$, in eV, for the V$_2$He complex. The
  migration paths are labelled by the initial and final
  configurations, as defined in \reffig{SubSubFig}. }.
\end{table}

The migration energies lie approximately 0.2 eV higher than those for
the corresponding single vacancy migration in afmD and afmI
Fe\cite{KlaverFeNiCr} and in Ni, where we found a vacancy migration
energy of 1.06 eV, in good agreement with other DFT
calculations\cite{DomainPerfect,Tucker10} and with the experimental
average value\cite{Erhart91} of $1.04 \pm 0.04$. We suggest that this
results from the additional energy required to move the He atom from
its central position in the V$_2$He complex back towards the lattice
site during migration, as observed in all cases. We also contrast
these results with those for divacancy migration. In afmD Fe, afmI Fe
and Ni we find migration energies for the divacancy along the
1b$\rightarrow$1b path of 0.370, 0.221 and 0.473 eV, respectively,
which are significantly lower than those for the V$_2$He complex. In
this case the difference arises not only from the energy required to
move He to an on-lattice site during migration but also from its
hindrance of the migrating Fe atom.

Vacancy-He exchange at 2nn provides an alternative migration path for
substitutional He to that of V$_2$He migration. We found energy
barriers for 2nn exchange as low as 0.47 and 0.55 eV in afmD and afmI
Fe, respectively, and a value of 0.94 eV in Ni. While these results
are lower than the migration energies for V$_2$He, the repulsive
interactions between a vacancy and substitutional He at 2nn [see
  Section \ref{soluteVacancySection}] mean that the equilibrium
concentrations of such configurations will be significantly lower than
the V$_2$He concentration, resulting, we believe, in a much lower
contribution to total diffusion. While this does strengthen our
position that V$_2$He migration dominates substitutional He diffusion
by the vacancy mechanism, a model including all the relevant migration
paths is necessary to answer this question conclusively.

Using the results presented here and in Section
\ref{soluteDefectSection}, we evaluate the expressions in
\eqnsfour{EmDiss1}{EmDiss2}{EmVac1}{EmVac2} and present the results in
\reftab{SubHeMigTab}.

\begin{table}[htbp]
\begin{ruledtabular}
\begin{tabular}{cccc}
  & afmD Fe & afmI Fe & Ni \\
\hline
$E_\mathrm{A}^\mathrm{diss.}$, \eqn{EmDiss1}  & 0.599-0.788 & 0.853-0.902 & 1.405 \\
$E_\mathrm{A}^\mathrm{diss.}$, \eqn{EmDiss2} & 2.411-2.600 & 2.810-2.859 & 2.756 \\
$E_\mathrm{A}^\mathrm{vac.}$, \eqn{EmVac1} & 2.066-2.413 & 2.232-2.251 & 2.192 \\
$E_\mathrm{A}^\mathrm{vac.}$, \eqn{EmVac2} & 0.254-0.601 & 0.275-0.294 & 0.841 \\
\end{tabular}
\end{ruledtabular}
\caption{\label{SubHeMigTab}Activation energies for substitutional He
  migration, in eV, by the dissociative,
  $E_\mathrm{A}^\mathrm{diss.}$, and vacancy,
  $E_\mathrm{A}^\mathrm{vac.}$, mechanisms for thermal
  (\eqns{EmDiss1}{EmVac1}) and supersaturated (\eqns{EmDiss2}{EmVac2})
  vacancy concentrations. For afmD and afmI Fe we give the range of
  possible values corresponding to the distinct migration paths in
  these states.}
\end{table}

The results clearly differentiate between the two mechanisms and show
a strong correlation to corresponding results in bcc
Fe\cite{Fu0507}. When thermal vacancies dominate we predict that
diffusion will proceed predominantly by the dissociative mechanism. If
a supersaturation of vacancies exists then the vacancy mechanism
clearly has the lowest activation energy. However, the vacancy
concentration also plays a critical role in determining which
mechanism dominates through the distinct way it enters the expressions
for the diffusion coefficients. For sufficiently low but still
supersaturated vacancy concentrations the dissociative mechanism may
become dominant. This is, however, most likely to be the case at low
temperatures where diffusion by either mechanism is likely to be
negligible. As such, we suggest that vacancy mediated diffusion is the
most important mechanism in conditions of vacancy supersaturation.

For the case of Ni, Philipps and Sonnenberg\cite{Philipps82} find an
activation energy for He diffusion of $0.81 \pm 0.04$ eV from
isothermal, He-desorption spectrometry experiments. They attribute
this result to the diffusion of substitutional He by the dissociative
mechanism, hindered by thermal vacancies, from which they infer an
energy for dissociation of He of 2.4 eV. Our results agree that
substitutional He migration will proceed by the dissociative mechanism
in a thermal vacancy population. There is, however, a 0.6 eV
difference between our calculated activation energy
[\reftab{SubHeMigTab}] and experiment. We also find a dissociation
energy for He from the substitutional site of 2.756 eV, which is in
excess of the inferred experimental value. This large discrepancy
suggests that the inferred experimental mechanism may not be
correct. Ab initio calculations show that interstitial He atoms bind
strongly to one another in Ni\cite{DomainPerfect}. As discussed
earlier, just such a mechanism was responsible for the suppression of
interstitial He migration in W and may also explain the experimental
result in Ni.

\subsubsection{Interstitial C and N migration}

\begin{figure}
\includegraphics[width=0.7\columnwidth]{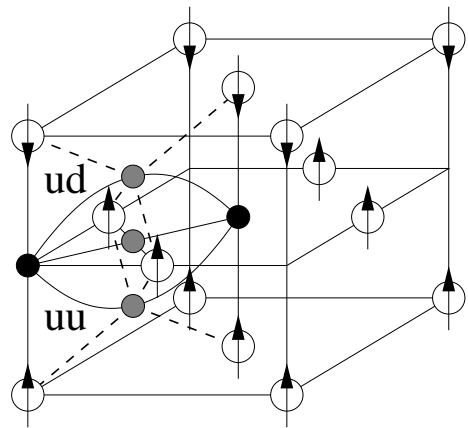}
\caption{\label{octaCandNMigFig}Possible migration paths for
  interstitial C and N in afmD Fe. Paths are shows for migration from
  initial to final octa sites (black circles) lying in the same
  magnetic plane via tetra ud, [110] crowdion and tetra uu
  intermediate sites (grey circles). Fe atoms (white circles) are
  shown with arrows to indicate the local moments. Migration between
  octa sites in adjacent magnetic planes have not been show for
  clarity. Coordinate axes are as in \reffig{soluteFig}.}
\end{figure}

The migration of interstitial C and N in both Fe and Ni goes from octa
site to adjacent octa site. In afmD Fe, there are three distinct
migration paths, depending on where the initial and final octa sites
lie. We label these as ``in-plane'', when the octa sites lie in the
same magnetic plane, ``uu'', when the octa sites lie in adjacent
magnetic planes with the same sign of magnetic moment and ``ud'', when
the octa sites lie in adjacent magnetic planes with the opposite sign
of magnetic moment. In afmI Fe, only the ``in plane'' and ``ud'' paths
are distinct and in Ni, all paths are equivalent. Each of these,
distinct, migration paths will be symmetrical about an intermediate
state lying in the plane that bisects the direct path between the two
octa sites. In what follows, we consider the tetra and $\langle
110\rangle$ crowdion sites as candidate intermediate states. Possible
migration paths for in-plane migration in afmD Fe are shown in
\reffig{octaCandNMigFig}, as an example.

For C, the results in \reftabs{solutesInIronTab}{solutesInNiTab} show
that the crowdion configurations are the lowest lying of the possible
intermediate states. We have performed NEB calculations in afmD Fe for
C migrating from the octa site to all of the distinct crowdion sites
in order to determine the energy profiles along these paths. We find a
single maximum in the energy at the crowdion configurations. We find
this is also the case for N migration via the crowdion configuration
in Ni, as will be discussed below. On this basis and given the
significant local dilatation necessary to form a crowdion, we make the
assumption that there will be a single energy maximum at all $\langle
110\rangle$ crowdion sites so that the MEPs and barrier heights for C
migration in afmD and afmI Fe and in Ni can be determined from the
data in \reftabs{solutesInIronTab}{solutesInNiTab}. The same can also
be said for N migration in afmD Fe along uu and ud paths. For all
other cases, however, the tetra sites are lower in energy and we have
performed NEB calculations with climbing image\cite{Henkelman00} to
investigate the migration energy profiles along these paths.

\begin{figure}
\includegraphics[width=\columnwidth]{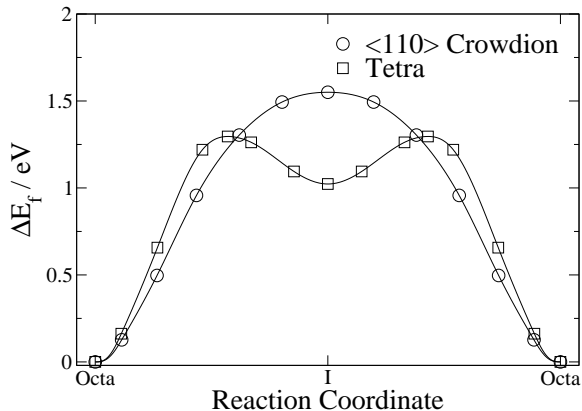}
\caption{\label{NmigNiFig}Formation energy difference,
  $\Delta\eform$, to the lowest energy octa configuration for N
  migration in Ni via tetra and $\langle 110\rangle$ crowdion
  intermediate states.}
\end{figure}

In afmD Fe, our calculations confirm that the tetra ud site is the
energy barrier for N migration. In afmI Fe, however, there is evidence
of a shallow minimum, 0.015 eV deep, around the tetra
configuration. Results in Ni, by contrast, show a clear double-peaked
structure in the energy profile. We present the results in
\reffig{NmigNiFig} and include the results for migration via the
crowdion site for comparison. The results show that the tetra N
configuration is a stable local minimum, with a depth of 0.273 eV
relative to the transition state, and not a saddle point, like the
crowdion configuration. Despite this, the MEP for N migration is still
via the tetra site.

\begin{table*}[htbp]
\begin{ruledtabular}
\begin{tabular}{ccccccccccccc}
\multirow{3}{*}{Path} & \multicolumn{6}{c}{C migration} & \multicolumn{6}{c}{N migration} \\
& \multicolumn{2}{c}{afmD Fe} & \multicolumn{2}{c}{afmI Fe} & \multicolumn{2}{c}{Ni} & \multicolumn{2}{c}{afmD Fe} & \multicolumn{2}{c}{afmI Fe} & \multicolumn{2}{c}{Ni} \\
& $\emig$ & TS/I & $\emig$ & TS/I & $\emig$ & TS/I & $\emig$ & TS/I & $\emig$ & TS/I & $\emig$ & TS/I \\
\hline
in plane & 2.033 & [110] crow. & 2.445 & [110] crow. & 1.628 & $\langle 110\rangle$ crow. & 1.558 & tetra ud & 1.899 & tetra$^\mathrm{*}$ & 1.296 & tetra$^\mathrm{*}$ \\
uu & 1.443 & [011] crow. uu & \multicolumn{2}{c}{as ud} & \multicolumn{2}{c}{as in plane} & 1.602 & [011] crow. uu & \multicolumn{2}{c}{as ud} & \multicolumn{2}{c}{as in plane} \\
ud & 1.310 & $[011]$ crow. ud & 2.113 & $[011]$ crow. ud & \multicolumn{2}{c}{as in plane} & 1.384 & $[011]$ crow. ud & 1.899 & tetra$^\mathrm{*}$ & \multicolumn{2}{c}{as in plane} \\
\end{tabular}
\end{ruledtabular}
\caption{\label{octaCandNMigTab}Migration energy barrier heights,
  $\emig$, in eV for interstitial C and N migration in afmD and afmI
  Fe and in Ni. Migration is between adjacent octa sites via a
  transition state/intermediate (TS/I), which is specified in the
  table, along all of the distinct paths for each particular reference
  state. Where the transition state/intermediate is only an
  intermediate state on the migration path, its name has been marked
  with an asterisk.}
\end{table*}

We summarise our results for the energy barriers and MEPs for
interstitial C and N migration in \reftab{octaCandNMigTab}. In the Fe
reference states there is a significant spread in the migration
barrier heights for C migration, both along distinct migration paths
and between the two states. In-plane migration clearly exhibits a
higher energy barrier in both states, which results directly from the
tetragonal distortion of the lattice and the subsequently higher
energy necessary to form the [110] crowdion transition state. The data
also suggests a significant dependence on the local magnetic order,
just as was seen for interstitial He migration. The large spread in
barrier heights means we cannot make any definitive predictions,
except that diffusion is three-dimensional. However, in any
thermodynamic average, the lower-energy paths will dominate, which
suggests an effective barrier height around 1.4 eV in afmD Fe and 2.1
eV in afmI Fe. The afmD Fe value is reasonably consistent with the
experimentally determined activation energies for C migration in
austenite of 1.626 eV\cite{Smith64} and 1.531 eV\cite{Agren86}.

In Ni, we find that C migrates via the crowdion site with an energy
barrier height of 1.63 eV. This contrasts with the 32 atom cell, where
the tetra pathway is preferred\cite{Siegel03}. Once again, this
demonstrates the inadequacy of using a 32 atom cell for solute
migration in fcc Fe and Ni. Experimental results, using a variety of
techniques applied both above and below the Curie temperature,
$T_\mathrm{C} = 627$ K, for Ni, yield activation energies in the range
1.43 to 1.75 eV\cite{LeClaireLB}, consistent with our results. The
experimental results also suggest that the influence of magnetism on
the migration barrier (and enthalpy of solution) for C is no more than
0.2 eV and suggests this is the likely error in using fm Ni results to
estimate those in the paramagnetic state.

Experimental results for C in Fe-Ni austenitic alloys, as discussed by
Thibaux {\it et al.}\cite{Thibaux07}, show only slight changes in C
mobility as a function of Ni composition in the range from 20 to 100
wt\% Ni. They also report an activation energy of 1.30 eV in an
Fe-31wt\%Ni austenitic alloy. Overall, our results, in conjunction
with the experimental results we have discussed, suggest that the
migration energy barrier for C migration will lie in the range 1.5$\pm
0.2$ eV across the whole composition range for Fe-Ni austenitic
alloys.

For N, the migration barrier lies between 1.38 and 1.60 eV in afmD Fe,
with a value of 1.90 eV in afmI Fe. As with C, the afmD Fe results
are, on average, lower than those for the afmI state. The result of an
Arrhenius fit to combined experimental diffusion data for N migration
in austenite gave a similar value of 1.75 eV\cite{LeClaireLB}. In Ni,
we find a barrier height of 1.30 eV, which is in excess of the
experimental activation energy reported by Lappalainnen and
Anttila\cite{Lappalainen87} of $0.99 \pm 0.12$ eV. In light of the
significant variation in experimental results for C migration in Ni,
these two results are in reasonable agreement and certainly to within
the 0.2 eV we have suggested earlier as a likely error when using
ferromagnetic Ni to model the paramagnetic state. Overall, these
results show that N migrates with a significantly lower barrier in Ni
than in austenitic Fe and we would expect to find an intermediate
value in Fe-Ni based alloys, more generally.

\subsection{Solute-solute interactions}

We have performed calculations to investigate the interactions between
pairs of He atoms in substitutional and tetra sites in afmD and afmI
Fe. Configurations with single substitutional and tetra-sited He atoms
at up to 2nn separation were found to consistently relax to a vacancy
containing two He atoms. While this does not yield any useful binding
energy data it does indicate that there is little or no barrier for
this process and places a lower limit on the capture radius of a
substitutional He of around 3 $\angs$. Results for pairs of
interacting substitutional and tetra-sited He atoms, as identified in
\reffigs{SubSubFig}{TetraTetraFig}, respectively, are given in
\reftab{HeHeTab}.

\begin{figure}
\includegraphics[width=0.7\columnwidth]{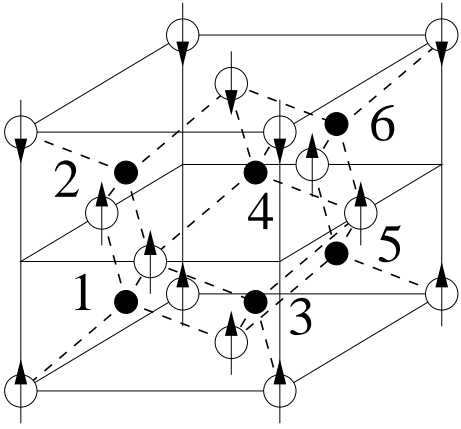}
\caption{\label{TetraTetraFig}Configurations for interacting
  tetra-sited solutes in afmD Fe. Configurations are labelled by the
  indices attached to the appropriate solute atom positions, shown in
  black. Fe atoms are shown in white with arrows to indicate the local
  moments. Coordinate axes are as in \reffig{soluteFig}.}
\end{figure}

\begin{table}[htbp]
\begin{ruledtabular}
\begin{tabular}{ccccc}
 & \multicolumn{2}{c}{afmD Fe} & \multicolumn{2}{c}{afmI Fe} \\ 
A-B/Config. & $\eform$ & $\ebind$ & $\eform$ & $\ebind$ \\ 
\hline 
S-S/1a & 7.115 & 0.934 & 7.423 & 0.946 \\ 
S-S/1b & 7.112 & 0.937 & \multicolumn{2}{c}{as S-S/1c} \\ 
S-S/1c & 6.976 & 1.073 & 7.419 & 0.950 \\ 
S-S/2a & 8.197 & -0.149 & 8.570 & -0.201 \\ 
S-S/2b & 8.109 & -0.060 & 8.493 & -0.123 \\ 
\hline 
T-T/1-2 & 8.614 & 0.313 & 9.692 & 0.242 \\ 
T-T/1-3 & 8.831 & 0.096 & \multicolumn{2}{c}{as T-T/2-4} \\ 
T-T/2-4 & 8.215 & 0.712 & 9.463 & 0.472 \\ 
T-T/1-4 & \multicolumn{2}{c}{rlx T-T/2-5} & 9.403 & 0.531 \\ 
T-T/1-5 & 8.700 & 0.227 & \multicolumn{2}{c}{as T-T/2-6} \\ 
T-T/2-6 & 8.643 & 0.284 & 9.428 & 0.506 \\ 
T-T/2-5 & 8.487 & 0.440 & 9.340 & 0.594 \\
\end{tabular}
\end{ruledtabular}
\caption{\label{HeHeTab} Formation and binding energies in eV for
  interacting pairs of He atoms in substitutional (S) and tetra (T)
  sites in afmD and afmI Fe. The configurations are labelled as in
  \reffigs{SubSubFig}{TetraTetraFig} for S-S and T-T pairs,
  respectively. In the afmD reference state the binding energies
  between tetra-sited pairs of He atoms have been calculated relative
  to two isolated tetra ud He. For interacting pairs of tetra-sited He
  atoms at 2nn and 3nn separation the configurations are labelled by
  the initial He positions, which due to the significant displacements
  under relaxation should not be taken as the final positions. Eshelby
  corrections for S-S pairs were found to be negligible but were -0.09
  eV for T-T pairs with a resulting increase in the T-T binding
  energies of up to 0.05 eV.}
\end{table}

Substitutional He pairs show consistent results across the Fe
reference states with a strong positive binding at 1nn and slightly
repulsive interactions at 2nn [\reftab{HeHeTab}]. In our calculations,
He atoms at 1nn relax directly towards one another by between 0.38 and
0.44~$\angs$, resulting in a consistent He-He separation of between
1.67 and 1.69~$\angs$. While still close to the lattice sites, these
displacements are in stark contrast to the insignificant displacements
observed at 2nn. Substitutional He pairs in Ni are similar: At 1nn the
He atoms are displaced towards one another by 0.37 $\angs$ to a He-He
separation of 1.75 $\angs$. The resultant binding energy, at 0.657 eV,
is less than in Fe but is in similar proportion to the substitutional
He to vacancy binding energy [see Section
  \ref{soluteVacancySection}]. At 2nn He remains on-lattice with a
repulsive binding energy of -0.16 eV.

Pairs of tetra-sited He atoms exhibit significant binding energies of
up to 0.7 eV in afmD Fe and 0.6 eV in afmI Fe. Such strong
interactions are consistently observed in bcc and fcc metals. Previous
ab initio calculations found binding energies of 0.47 eV in
Ni\cite{DomainPerfect}, 0.7 eV in Pd\cite{Zeng09} and Al\cite{Yang},
0.4 eV in bcc Fe\cite{Fu0507} and 1.0 eV in W\cite{Becquart06}. At 1nn
separation, the He atoms in afmD and afmI Fe are displaced from the
tetra sites only slightly under relaxation. The resulting He-He
``bonds'' all lie along one of the axes of the unit cell with lengths
in a small range from 1.62 to 1.68~$\angs$, which is consistent with
those found for substitutional He pairs at 1nn. At 2nn and 3nn the He
atoms displace significantly towards one another under relaxation from
the initial tetra sites, resulting in He-He separations from 1.51 to
1.65~$\angs$. These displacements are sufficiently large to take the
He atoms either to the edge of their initial tetrahedral regions or
into the adjacent octahedral region via one of the faces of the
tetrahedron. This is most pronounced for the 3nn T-T/2-5
configuration, which in afmI Fe relaxed to a configuration with the He
atoms within the octahedral region and symmetrically opposite the
central position along the [111] axis. The situation is similar for
the afmD state but one He atom is significantly closer to the central
position. It is worth noting that this is the most stable
configuration in afmI Fe and the second most stable in afmD Fe. The
large binding energies result, simply, from the co-operative
dilatation of the lattice and the reduction of repulsive He-Fe
interactions, which are naturally greatest when the two He atoms are
in close proximity. The results at 2nn and 3nn separations show that
the local dilatation of the lattice around a single interstitial He
results in an attractive force to other interstitial He atoms up to at
least 3 $\angs$ away and encourages the formation of clusters.

To investigate interstitial cluster formation further we have
determined the most stable configurations with three and four He atoms
in afmD and afmI Fe. For a fixed number of He atoms we found many
distinct configurations with similar energies but the most stable
clusters were reasonably predictable from a simple pair interaction
model, given the data in \reftab{HeHeTab}. In afmD Fe, the most stable
He$_3$ configuration found had two He atoms in a 2-4 formation [see
  \reffig{TetraTetraFig}] with the third occupying the nearest octa
site. In afmI Fe, the most stable was an L-shaped 1-2-3 cluster. In
the most stable He$_4$ clusters, all He atoms occupied tetra sites in
a rectangular-planar formation with 1nn edges, such as a 1-2-3-4
cluster. This is, in fact, the most stable arrangement found in afmI
Fe, whereas in afmD Fe a square-planar configuration with all He atoms
in tetra ud sites was the most stable. The total binding energies for
the most stable clusters are given in \reftab{IntHeClusterTab} along
with results in Ni\cite{DomainPerfect}.

\begin{table}[htbp]
\begin{ruledtabular}
\begin{tabular}{cccc}
Cluster & afmD Fe & afmI Fe & Ni \\
\hline
He$_2$ & 0.712 & 0.594 & 0.47 \\
He$_3$ & 1.537 & 1.374 & 1.25 \\
He$_4$ & 2.637 & 2.561 & \\
\end{tabular}
\end{ruledtabular}
\caption{\label{IntHeClusterTab} Total binding energies, in eV, for
  the most stable interstitial He clusters containing up to 4 He
  atoms. Results in Ni are from the work of Domain and
  Becquart\cite{DomainPerfect}. Eshelby corrections were found to be
  -0.19 and -0.34 eV for He$_3$ and He$_4$ clusters, respectively,
  with corresponding increases in the binding energies of 0.14 and
  0.27 eV.}
\end{table}

The strong clustering tendency of interstitial He is clearly
demonstrated by the data. Application of the Eshelby corrections only
enhances this effect. The binding energy for an additional He, that
is, $\ebind(\mathrm{He}_n) - \ebind(\mathrm{He}_{n-1})$, increases
with $n$ for the small clusters studied here. We would expect this,
however, to plateau to an additional binding energy of around 1 eV per
He atom in afmD and afmI Fe and in Ni, given that the co-operative
dilatation of the lattice that gives rise to the binding happens
locally. Such strong clustering can, not only, result in an effective
reduction in interstitial He mobility as He concentration increases
but is also a critical first step in the spontaneous formation of
Frenkel-pair defects, as observed in gold\cite{Thomas81}. Indeed, the
most stable He$_4$ configurations found here show a significant
displacement of the nearest Fe atom to the cluster off-lattice by 0.94
$\angs$ in afmD Fe and 1.36 $\angs$ in afmI Fe.  We consider this
possibility further in Section \ref{soluteVacancyClusteringSection} in
the context of V$_m$He$_n$ clustering.

\begin{table}[htbp]
\begin{ruledtabular}
\begin{tabular}{ccccc}
 & \multicolumn{2}{c}{afmD Fe} & \multicolumn{2}{c}{afmI Fe} \\
A-B/Config. & $\eform$ & $\ebind$ & $\eform$ & $\ebind$ \\
\hline
C-C/1a & -17.490 & -0.104 & -17.572 & -0.141 \\
C-C/1b & -17.487 & -0.106 & \multicolumn{2}{c}{N/A} \\
C-C/1c & -17.559 & -0.034 & -17.561 & -0.151 \\
C-C/2a & -17.420 & -0.174 & -17.487 & -0.226 \\
C-C/2b & -17.617 & 0.023 & -17.655 & -0.058 \\
\hline
N-N/1a & -17.010 & -0.195 & -17.037 & -0.204 \\
N-N/1b & -17.035 & -0.170 & \multicolumn{2}{c}{N/A} \\
N-N/1c & -17.131 & -0.074 & -17.020 & -0.221 \\
N-N/2a & -17.054 & -0.150 & -17.075 & -0.167 \\
N-N/2b & -17.212 & 0.008 & -17.172 & -0.070 \\
\end{tabular}
\end{ruledtabular}
\caption{\label{octaoctaTab} Formation and binding energies, in eV,
  for interacting pairs of octa-sited C and N interstitials. The
  configurations are as labelled in \reffig{octaoctaFig}. Eshelby
  corrections were -0.03 and -0.06 eV in afmD and afmI Fe,
  respectively, with resultant increases in the binding energy of 0.02
  and 0.03 eV.}
\end{table}

\begin{figure}
\includegraphics[width=0.7\columnwidth]{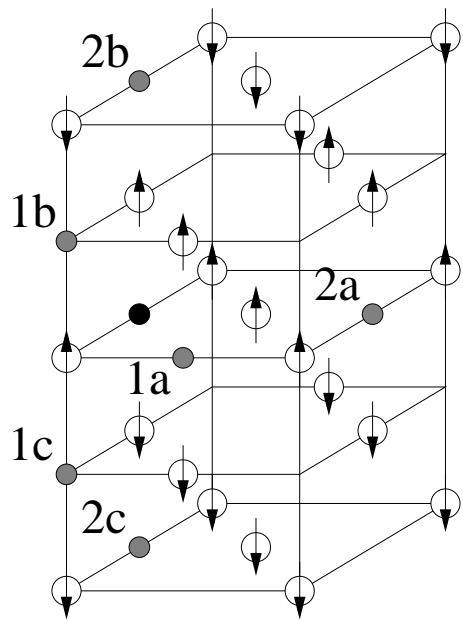}
\caption{\label{octaoctaFig}Configurations for A-B pairs of
  interacting octa-sited interstitials in afmD Fe. Species A is shown
  in black and species B in grey along with the configuration
  labels. Fe atoms are shown in white with arrows to indicate the
  local moments. The lowest symmetry afmD state is shown to uniquely
  identify all of the distinct configurations. Some of these
  configurations will be symmetry equivalent in the afmI state. Coordinate axes
  are as in \reffig{soluteFig}.}
\end{figure}

Interactions between pairs of octa-sited C and N atoms at up to 2nn
separation in afmD and afmI Fe are given in \reftab{octaoctaTab}. The
interactions are, generally, repulsive at both 1nn and 2nn, with a
reasonable consistency between the two reference states, although the
repulsion is slightly stronger in the afmI state. For C, the pair
binding energy is between -0.1 and -0.15 eV at 1nn and more repulsive
at 2nn at up to around -0.2 eV. By contrast, N pairs exhibit stronger
repulsion than C at 1nn, at around -0.2 eV and weaker repulsion at
2nn, where the binding energy is at most around -0.15 eV. Calculations
for C-C pairs at up to 4nn separation in afmI Fe found a maximal
binding energy 0.02 eV. We conclude that there will be no appreciable
positive binding of C-C and N-N pairs in our reference states for
austenite.

Experimental determinations of C-C and N-N interaction energies in
austenite are discussed in a review by Bhadeshia\cite{Bhadeshia04}. If
quasichemical theory, which only includes 1nn interactions, is used to
interpret the existing thermodynamic data then a pair binding energy
of -0.09 eV is found for C and -0.04 eV for N. Our results for C in
afmD and afmI Fe are in good agreement with this value and while we do
find a repulsive interaction between N-N pairs, we find a stronger
repulsion than for C, which is contrary to the results of this
analysis. A more detailed analysis can be performed using M\"ossbauer
spectroscopy data to study the distribution of C atoms in the Fe
matrix, which can be compared with the results of Monte Carlo
simulations to determine the solute interaction energies at 1nn and
2nn\cite{Oda94}. Such an analysis yields C-C binding energies of -0.04
eV at 1nn and less than -0.08 eV at 2nn and N-N binding energies of
-0.08 eV and -0.01 eV at 1nn and 2nn, respectively\cite{Oda94}. Our
results follow the same pattern for the relative strengths of
repulsion but are significantly in excess of the results of this
analysis. The agreement is still impressive, however, given the level
of extrapolation between our two ordered magnetic states at 0K and
temperatures where paramagnetic austenite is stable.

For C in fm Ni, Siegel and Hamilton\cite{Siegel03} found C-C binding
energies at 1nn and 2nn of 0.01 and -0.01 eV, respectively, using
comparative DFT calculations to those performed in this work. They,
furthermore, show that this negligible level of binding is consistent
with the experimental estimates of the C-C pair concentration as a
function of total C concentration\cite{Numakura00}.

From the data presented above we would suggest that C-C and N-N
interactions in Fe-based austenitic alloys will be repulsive at 1nn
and 2nn, with binding energies in the range from -0.1 to -0.2 eV. We
would, furthermore, expect the level of repulsion to be reduced as a
function of increasing Ni concentration. 

\subsection{Interactions with substitutional Ni and Cr solutes in Fe}

As an initial step in the investigation of the interactions of He, C
and N with substitutional Ni and Cr solutes in austenite we have
calculated the formation energies for single substitutional Ni and Cr
and present the results in \reftab{subNiCrTab}.\footnote{These results
  agree to within errors with our previous work\cite{KlaverFeNiCr} at
  a lower plane-wave cutoff energy and direct the reader there for
  further discussion.} On this basis, the results of our calculations
of the interactions between He, C and N solutes and substitutional Ni
and Cr solutes in afmD and afmI Fe are presented in
\reftab{subNiCrHeTab}.

\begin{table}[htbp]
\begin{ruledtabular}
\begin{tabular}{ccccc}
Config. & \multicolumn{2}{c}{fct afmD} & \multicolumn{2}{c}{fct afmI} \\
& $\eform$ & $\mu$ & $\eform$ & $\mu$ \\
\hline
Sub. Ni & 0.083 & 0.039 & 0.145 & -0.301 \\
Sub. Cr & 0.263 & 0.843 & 0.061 & 1.120 \\
\end{tabular}
\end{ruledtabular}
\caption{\label{subNiCrTab} Formation energies, $\eform$, in eV and
magnetic moments, $\mu$, in $\mu_B$ for substitutional Ni and Cr
solutes in austenitic Fe. The sign of the moments indicates whether there
is alignment (positive) or anti-alignment (negative) with the
moments of the atoms in the same magnetic plane.}
\end{table}

\begin{table}[htbp]
\begin{ruledtabular}
\begin{tabular}{ccccc}
 & \multicolumn{2}{c}{afmD Fe} & \multicolumn{2}{c}{afmI Fe} \\
A-B/cfg & $\eform$ & $\ebind$ & $\eform$ & $\ebind$ \\
\hline
sub Ni-sub He/1a & 4.032 & 0.076 & 4.212 & 0.117 \\
sub Ni-sub He/1b & 4.035 & 0.073 & \multicolumn{2}{c}{as 1c} \\
sub Ni-sub He/1c & 4.018 & 0.090 & 4.233 & 0.097 \\
\hline
sub Ni-tetra He/1b & 4.496 & 0.051 & 4.979 & 0.133 \\
sub Ni-tetra He/2a & 4.500 & 0.047 & 5.078 & 0.034 \\
sub Ni-tetra He/2d & 4.480 & 0.067 & 5.062 & 0.050 \\
\hline
sub Ni-octa C/1a & -8.692 & -0.021 & -8.717 & 0.006 \\
sub Ni-octa C/1b & -8.673 & -0.040 & \multicolumn{2}{c}{as 1c} \\
sub Ni-octa C/1c & -8.729 &  0.016 & -8.643 & -0.069 \\
\hline
sub Ni-octa N/1a & -8.439 & -0.080 & -8.417 & -0.058 \\
sub Ni-octa N/1b & -8.432 & -0.087 & \multicolumn{2}{c}{as 1c} \\
sub Ni-octa N/1c & -8.445 & -0.074 & -8.349 & -0.126 \\
\hline
sub Cr-sub He/1a & 4.353 & -0.065 & 4.284 & -0.038 \\
sub Cr-sub He/1b & 4.358 & -0.070 & \multicolumn{2}{c}{as 1c} \\
sub Cr-sub He/1c & 4.433 & -0.145 & 4.341 & -0.095 \\
\hline
sub Cr-tetra He/1b & 4.609 & 0.118 & 4.883 & 0.145 \\
sub Cr-tetra He/2a & 4.746 & -0.019 & 5.005 & 0.023 \\
sub Cr-tetra He/2d & 4.781 & -0.054 & 5.024 & 0.004 \\
\hline
sub Cr-octa C/1a & -8.647 & 0.114 & -8.845 & 0.050 \\
sub Cr-octa C/1b & -8.628 & 0.094 & \multicolumn{2}{c}{as 1c} \\
sub Cr-octa C/1c & -8.730 & 0.197 & -8.826 & 0.030 \\
\hline
sub Cr-octa N/1a & -8.597 & 0.258 & -8.729 & 0.169 \\
sub Cr-octa N/1b & -8.566 & 0.227 & \multicolumn{2}{c}{as 1c} \\
sub Cr-octa N/1c & -8.574 & 0.235 & -8.704 & 0.145 \\
\end{tabular}
\end{ruledtabular}
\caption{\label{subNiCrHeTab} Formation and binding energies, in eV,
  for substitutional Ni/Cr (species A) to substitutional He, tetra He
  and octa C/N (species B) with configurations labelled as in
  \reffigss{SubSubFig}{SubTetraFig}{SubOctaFig}, respectively. Eshelby
  corrections to $\eform$ were found to be -0.02 eV when interstitial
  solutes were present but negligible for all other quantities. }
\end{table}

\begin{figure}
\includegraphics[width=0.7\columnwidth]{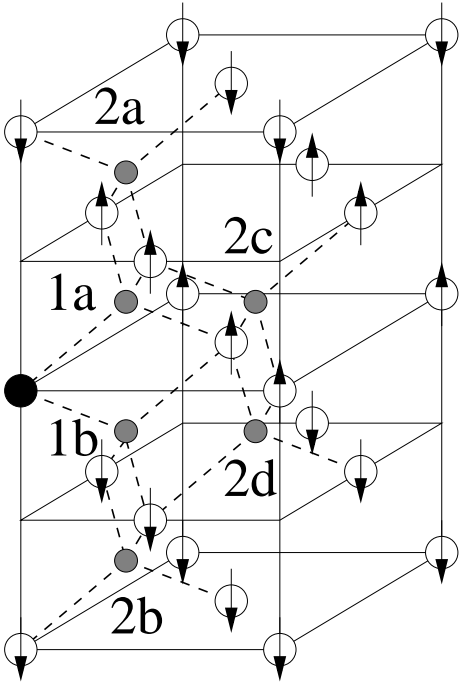}
\caption{\label{SubTetraFig}Configurations for interactions between a
  substitutionally sited species (A) and a tetrahedrally sited species
  (B) in afmD Fe. Species A is shown in black, and species B in
  grey. Configurations are labelled by the position of the
  tetrahedrally sited species, as shown. Fe atoms are shown in white
  with arrows to indicate the local moments. Coordinate axes are as in
  \reffig{soluteFig}.}
\end{figure}

\begin{figure}
\includegraphics[width=0.8\columnwidth]{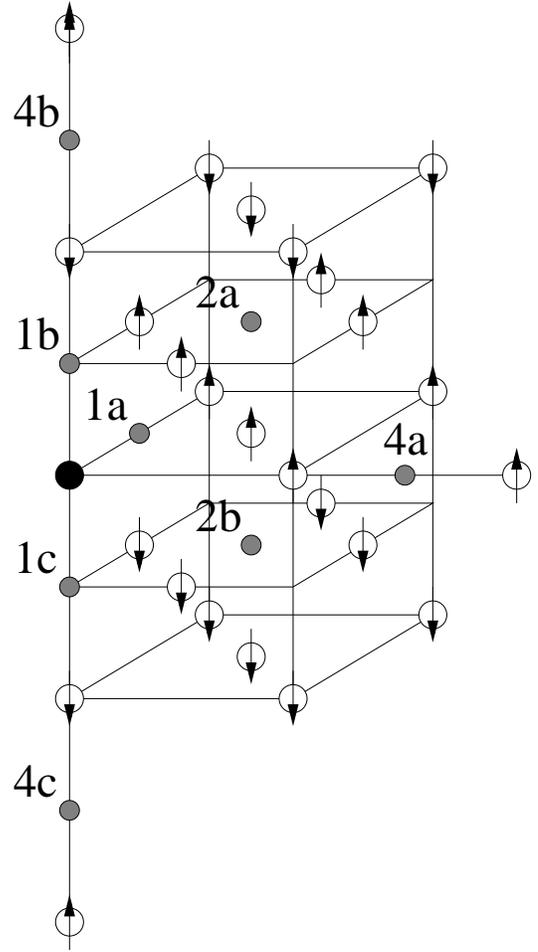}
\caption{\label{SubOctaFig}Configurations for interactions between a
  substitutionally sited species A and octa-sited species B in the
  fcc/fct afmD reference state. Species A is shown in black and
  species B in grey along with the configuration labels. Fe atoms are
  shown in white with arrows to indicate the local moments. The lowest
  symmetry afmD state is shown to uniquely identify all of the
  distinct configurations. Some of these configurations will be
  symmetry equivalent in the afmI state. Coordinate axes are as in
  \reffig{soluteFig}.}
\end{figure}

In both Fe reference states, substitutional He binds weakly to Ni, by
around 0.1 eV, and has a repulsive interaction with Cr of -0.1 eV. The
similarity to vacancy-substitutional Ni/Cr binding is
striking\cite{KlaverFeNiCr}. The similarity also extends to the local
moments on 1nn atoms surrounding the substitutional He and vacancy, as
was discussed in Section \ref{solutesInIron}. These results are also
consistent with Ni and Cr acting as slightly oversized and undersized
solutes, respectively, when interacting with point defects in afmD and
afmI Fe, as discussed previously\cite{KlaverFeNiCr}. We would expect
the interactions of other transition metal solutes with substitutional
He to be readily inferred from their interactions with vacancies.

Interstitial He binds weakly to Ni by, on average, 0.09 eV at 1nn and
0.05 eV at 2nn in the Fe reference states. We also observe weak
positive binding with Cr, but only at 1nn, where the binding energy
is, on average, 0.13 eV. Closer observations of the configurations
revealed that He relaxed slightly away from Ni, but toward Cr at
1nn. Ni also remained closer to the lattice site than Cr. These
geometrical results are consistent with Ni and Cr behaving as
oversized and undersized solutes, respectively, despite both
exhibiting binding to interstitial He, although the binding to Cr is
marginally greater. The level of binding suggests that Ni and Cr may
act as weak traps for migrating interstitial He at low
concentrations. However, with increasing concentration and, therefore,
likelihood that He remains in similar local environments as it
migrates, a direct study of the local composition dependence of the
migration energy becomes necessary. From the binding energy data we
can speculate, however, that such a dependence will also be weak and
maintain our earlier suggestion that the activation energy for
interstitial He migration will lie in the 0.1 to 0.2 eV range in
concentrated Fe-Cr-Ni alloys.

The interactions of octa C and N with substitutional Ni and Cr are
reasonably consistent in both afmD and afmI Fe. For C, interactions
with Ni are minimal, although slightly repulsive, at 1nn, whereas
positive binding is observed with Cr on the order of 0.1 eV. The
interactions of N are similar to those of C but significantly stronger
and exhibit a repulsion of around -0.1 eV to Ni and attraction to Cr
of around 0.2 eV. The repulsive interactions with Ni are consistent
with the lower solubility of C and, particularly, N in fcc Ni,
compared to afmD and afmI Fe [see
  \reftabs{solutesInIronTab}{solutesInNiTab}], and suggests that the
interactions are cumulative. In the case of Cr, such cumulative
interactions would encourage the formation of Cr-C/N complexes and the
precipitation of Cr-carbonitrides, as observed experimentally in
non-stabilised austenitic stainless steels\cite{Sourmail01}, under
conditions where these elements are mobile, that is at high
temperatures or in irradiated environments.

\section{Solute interactions with point defects}
\label{soluteDefectSection}

In this section we consider the interactions of He, C and N with a
single vacancy (V), in small vacancy-solute clusters, V$_m$X$_n$, and
with the [001] self-interstitial (SI) dumbbell in afmD and afmI Fe and
in Ni. We present the formation (and binding) energies of the
underlying and most stable defects and defect clusters in
\reftab{PointDefectTab}, as found previously\cite{KlaverFeNiCr}. Pairs
of vacancies were consistently most stable at 1nn separation. The most
stable tetra-vacancy cluster consists of a tetrahedral arrangement of
vacancies at 1nn to each other. The most stable tri-vacancy cluster is
formed from this by placing an atom near the tetra-vacancy
centre. Finally, the most stable hexa-vacancy is an octahedral
arrangement of vacancies with 1nn edges.

\begin{table}[htbp]
\begin{ruledtabular}
\begin{tabular}{cccc}
Defect & afmD Fe & afmI Fe & Ni \\
\hline
vacancy & 1.812 & 1.957 & 1.352 \\
di-vacancy & 3.443 & 3.840 & 2.688 \\
           & (0.181) & (0.075) & (0.016) \\
tri-vacancy & 4.790 & 5.285 & \\
            & (0.646) & (0.587) & \\
tetra-vacancy & 6.479 & 7.097 & 4.956 \\
              & (0.768) & (0.733) & (0.451) \\
hexa-vacancy & 8.378 & 9.210 & 6.865 \\
             & (2.493) & (2.534) & (1.245) \\
$[001]$ SI dumbbell & 3.196 & 3.647 & 4.135 \\
\end{tabular}
\end{ruledtabular}
\caption{\label{PointDefectTab} Formation energies, $\eform$, in eV,
  for the vacancy, the most stable di-, tri-, tetra- and hexa-vacancy
  clusters, as found by Klaver {\it et al.}\cite{KlaverFeNiCr} and the
  [001] SI dumbbell in afmD and afmI Fe and in Ni. Total binding
  energies, $\ebind$, in eV, are given for the vacancy clusters in
  brackets below the formation energies. The results in Fe are
  consistent with those found previously\cite{KlaverFeNiCr}. Results
  in Ni compare well to other DFT
  calculations\cite{Megchiche06,DomainPerfect,Zu,Megchiche10,Tucker10}. Eshelby
  corrections were found to be negligible except for the tetra-vacancy
  in Ni at -0.03 eV, the hexa-vacancy at -0.06 -0.03 and -0.05 eV in
  afmD Fe, afmI Fe and Ni, respectively, and the dumbbell at -0.05,
  -0.08 and -0.10 eV in afmD Fe, afmI Fe and Ni, respectively. The
  only non-negligible effect on binding energies was for the
  hexa-vacancy, where increases of 0.05 eV, 0.03 and 0.03 eV apply in
  afmD Fe, afmI Fe and Ni, respectively.}
\end{table}

\subsection{Vacancy - solute interactions}
\label{soluteVacancySection}

\begin{table}[htbp]
\begin{ruledtabular}
\begin{tabular}{ccccc}
 & \multicolumn{2}{c}{afmD Fe} & \multicolumn{2}{c}{afmI Fe} \\
A-B/Config. & $\eform$ & $\ebind$ & $\eform$ & $\ebind$ \\
\hline
V-sub He/1a & 5.216 & 0.620 & 5.519 & 0.623 \\
V-sub He/1b & 5.221 & 0.615 & \multicolumn{2}{c}{as V-sub He/1c} \\
V-sub He/1c & 5.180 & 0.656 & 5.538 & 0.604 \\
V-sub He/2a & 5.940 & -0.103 & 6.293 & -0.151 \\
V-sub He/2b & 5.854 & -0.018 & \multicolumn{2}{c}{as V-sub He/2c} \\
V-sub He/2c & 5.919 & -0.083 & 6.253 & -0.111 \\
\hline
V-tetra He/Sub. & 4.024 & 2.251 & 4.185 & 2.740 \\
V-tetra He/2a & 6.328 & -0.053 & \multicolumn{2}{c}{as V-tetra He/2b} \\
V-tetra He/2b & 6.408 & -0.133 & 6.932 & -0.008 \\
V-tetra He/2c & 6.377 & -0.101 & \multicolumn{2}{c}{as V-tetra He/2d} \\
V-tetra He/2d & 6.233 & 0.042 & 6.883 & 0.041 \\
\hline
V-C/Sub. & -6.981 & -0.004 & -6.244 & -0.655 \\
V-C/1a & -7.165 & 0.180 & -7.276 & 0.377 \\
V-C/1b & -7.040 & 0.056 & \multicolumn{2}{c}{as V-C/1c} \\
V-C/1c & -7.268 & 0.283 & -7.186 & 0.287 \\
V-C/2a & -7.031 & 0.046 & \multicolumn{2}{c}{as V-C/2b} \\
V-C/2b & -7.018 & 0.033 & -6.948 & 0.049 \\
\hline
V-N/Sub. & \multicolumn{2}{c}{unstable} & -5.153 & -1.510 \\
V-N/1a & -7.230 & 0.440 & -7.275 & 0.612 \\
V-N/1b & -7.065 & 0.275 & \multicolumn{2}{c}{as V-N/1c} \\
V-N/1c & -7.217 & 0.427 & -7.161 & 0.498 \\
V-N/2a & -6.887 & 0.097 & \multicolumn{2}{c}{as V-N/2b} \\
V-N/2b & -6.883 & 0.092 & -6.769 & 0.106 \\
\end{tabular}
\end{ruledtabular}
\caption{\label{vacsolTab} Formation and binding energies in eV for
  vacancy (species A) to substitutional He, tetra He and octa C/N
  (species B) with configurations labelled as in
  \reffigss{SubSubFig}{SubTetraFig}{SubOctaFig},
  respectively. Configurations with a single solute atom in the
  substitutional position (Sub.), where stable, are also considered as
  an interstitial solute interacting with a vacancy. In afmD Fe, the
  vacancy-tetra He binding energies were calculated relative to tetra
  ud He. Binding energies between octa C and N solutes and a vacancy
  at 3nn and 4nn separations were investigated but did not exceed 0.03
  eV. The only non-negligible Eshelby corrections found were for the
  binding energies between a vacancy and interstitial solutes at no
  more than 0.02 eV in magnitude.}
\end{table}

We present the formation and binding energies for configurations
containing a single vacancy and solute atom, at up to 2nn separation,
in \reftab{vacsolTab}. 

\subsubsection{V-He binding}

We observe strong binding of between 0.60 and 0.66 eV for V-Sub He
pairs at 1nn in both Fe reference states. This is significantly
greater than the binding between vacancy pairs\cite{KlaverFeNiCr} and
represents the simplest case of enhanced vacancy binding by He, as we
will discuss in what follows. We find that He does not remain
on-lattice at 1nn to a vacancy but relaxes to a position, best
described as, at the centre of a divacancy. With this perspective, the
V-Sub He binding represents the significant energetic preference of He
for the greater free volume available at the centre of a divacancy
over a single vacancy. At 2nn, the interactions are repulsive, at
around -0.1 eV, which is slightly greater than that observed between
vacancy pairs\cite{KlaverFeNiCr}. He remains on-lattice in these
configurations, which explains the lack of enhanced binding at 2nn
separation. The situation in Ni is very similar, where we find binding
energies of 0.356 eV and -0.127 eV at 1nn and 2nn, respectively.

Interstitial He binds strongly to a vacancy to form a substitutional
He configuration. The same is also true in Ni, where we find a binding
energy of 2.627 eV, in good agreement with previous
work\cite{DomainPerfect}. Configurations with tetra He at 1nn to a
vacancy are unstable. At 2nn, however, we find stable configurations
with weak repulsive or attractive binding, depending on the
configuration. The fact that no stable configurations were found with
tetra He at up to 2nn from a substitutional He atom demonstrates that
the addition of a single He to a vacancy significantly increases the
capture radius for interstitial He. We expect this effect to increase
with the subsequent addition of He, given the additional pressure and
dilatation that would be exerted on the surrounding lattice.

\subsubsection{V-C and V-N binding}

C binds to a vacancy by up to 0.38 eV at 1nn in the Fe reference
states and weakly at 2nn. This level of binding agrees well with
previous experimental and theoretical work in austenite and austenitic
alloys\cite{Slane04}. We find that V-N binding is significantly
stronger than for C with binding energies in the range from 0.3 to 0.6
eV at 1nn and around 0.1 eV at 2nn. For both C and N, the
substitutional configuration is strongly disfavoured. As discussed in
Section \ref{solutesInIron}, the substitutional C and N configurations
in afmD Fe were found to be unstable and the configuration labelled
V-C/Sub in \reftab{vacsolTab} has C in a stable position off-lattice
by 0.77 $\angs$. Overall, these results bear a strong similarity to
those found in bcc Fe\cite{Domain04}, where binding energies of 0.47
and 0.71 eV were found for C and N at 1nn to a vacancy, respectively.

Results in Ni are broadly similar to those in Fe. We find a V-C
binding energy of 0.062 eV at 1nn and 0.121 eV at 2nn. V-N binding is,
again, stronger, than C, with energies of 0.362 eV and 0.165 eV at 1nn
and 2nn, respectively. We also find a strong repulsion from the
substitutional site. We note that the V-C binding at 1nn seems
anomalously low, given the other results but no problems were found
with this calculation and other test calculations found the same
stable structure and energy.

The significant V-C and V-N binding energies suggest that the
relatively less mobile solutes could act as vacancy traps, much as was
found in bcc Fe\cite{Vehanen82,Takaki83,Tapasa07,Fu08}. This would
certainly be the case if dissociation of the complex was required
before the vacancy could freely migrate but the possibility of
cooperative migration also exists. In the fcc lattice there are many
possible migration pathways that would avoid the dissociation of this
complex, including some that would maintain a 1nn separation. A
complete study of these possibilities is beyond the scope of this work
but preliminary calculations in Ni show that the energy barriers for C
and N jumps that would maintain a 1nn separation to the vacancy are
around half the value for the isolated solutes at around 0.75 eV. In
contrast, vacancy jumps that maintain a 1nn separation were found to
be significantly higher than those for the isolated vacancy but jumps
from 1nn to 2nn separation and back exhibited lower or comparable
energy barriers. Whilst these calculations are preliminary, they do
indicate the distinct possibility of cooperative vacancy-solute motion
that would avoid dissociation of the complex. The implications for an
absence of vacancy pinning and for the enhanced diffusion of C and N
solutes in the presence of vacancies in austenitic alloys makes this
an interesting subject for further study.

\subsection{Vacancy - solute clustering}
\label{soluteVacancyClusteringSection}

Small Vacancy-He (V$_m$He$_n$) clusters have been found to be highly
stable both experimentally\cite{Vassen91,Fedorov9698,Cao11,Ono11} and
using DFT
techniques\cite{Perfect,DomainPerfect,Yang,Zeng09,Fu0507,Becquart09}
in a number of metals and are, therefore, critically important as
nuclei for void formation. Experimental evidence in bcc
Fe\cite{Vehanen82,Takaki83} has also shown that C can act as a vacancy
trap through the formation of small, stable V$_m$C$_n$ clusters, which
has been confirmed in a number of DFT
studies\cite{Domain04,Forst06,Lau07,Fu08,Ortiz0709}. Small V$_m$N$_n$
clusters have also been shown to exhibit similar
stability\cite{Domain04}.

In this section we present the results of a large number of DFT
calculations to find the most stable V$_m$X$_n$ clusters, where X is
He, C or N, in afmD and afmI Fe. A comprehensive search for the most
stable configuration was only practicable for the smaller
clusters. For larger clusters, a number of distinct initial
configurations, based around the most stable smaller clusters, were
investigated to improve the likelihood that the most stable
arrangement was found. The total binding energies for the most stable
configurations can be found in \reftab{clusterBindingTab}.

\begin{table}[htbp]
\begin{ruledtabular}
\begin{tabular}{ccc|cccc}
\multirow{2}{*}{Cluster} & afmD Fe & afmI Fe&  \multirow{2}{*}{Cluster} & afmD Fe & afmI Fe & Ni \\
 & $\ebind$ & $\ebind$ & & $\ebind$ & $\ebind$ & $\ebind$ \\
\hline
VHe & 2.251 & 2.740 & VC & 0.283 & 0.377 & 0.121 \\
VHe$_2$ & 3.845 & 4.627 & VC$_2$ & 0.484 & 0.795 & 0.422 \\
VHe$_3$ & 5.674 & 6.588 & VC$_3$ & 0.423 & 0.484 & -0.206 \\
VHe$_4$ & 7.452 & 8.609 & V$_2$C & 0.499 & 0.550 & 0.211 \\
VHe$_5$ & 9.239 & 10.305 & V$_4$C & 1.107 & 1.307 & 0.718 \\
VHe$_6$ & 10.845 & 12.015 & V$_6$C & 3.546 & 3.253 & 1.531 \\
V$_2$He & 2.907 & 3.363 & VN & 0.440 & 0.612 & 0.362 \\
V$_2$He$_2$ & 5.575 & 6.430 & VN$_2$ & 0.981 & 1.295 & 0.872 \\
V$_2$He$_3$ & 7.791 & 8.990 & VN$_3$ & 1.264 & 1.341 & 0.877 \\
V$_2$He$_4$ & 10.197 & 11.682 & VN$_4$ & 1.371 & 1.514 & 0.933 \\
V$_3$He & 3.711 & 4.237 & VN$_5$ & 1.439 & 1.516 & 0.651 \\
V$_3$He$_2$ & 6.458 & 7.461 & VN$_6$ & 1.474 & 1.482 & 0.246 \\
V$_3$He$_3$ & 9.323 & 10.857 & V$_2$N & 0.743 & 0.933 & 0.558 \\
V$_3$He$_4$ & 11.750 & 13.685 & V$_4$N & 1.364 &1.573 & 1.047 \\
V$_4$He & 4.475 & 4.993 & V$_6$N & 3.224 & 3.466 & 2.033 \\
V$_4$He$_2$ & 7.504 & 8.542 & & & & \\ 
V$_4$He$_3$ & 10.565 & 12.120 & & & & \\
V$_4$He$_4$ & 13.606 & 15.711 & & & & \\
V$_6$He & 6.566 & 7.191 & & & & \\
\end{tabular}
\end{ruledtabular}
\caption{\label{clusterBindingTab} Total binding energies, $\ebind$,
  in eV for the most stable V$_m$X$_n$ clusters found in afmD and afmI
  Fe, where X is He, C or N. Results in Ni are also given for C and N
  and can be found in the literature for
  He\cite{Perfect,DomainPerfect}. Eshelby corrections to $\ebind$ for
  V$_m$He$_n$ clusters were found to be below 0.05 eV in magnitude
  except for VHe$_5$ and V$_m$He$_n$ with $m$ and $n$ equal to 3 or 4,
  which were below 0.1 eV and VHe$_6$, which was 0.2 eV. For C and N
  clusters, the corrections were below 0.02 eV in magnitude except for
  those with six vacancies or with four or more N atoms, where the
  corrections were up to 0.1 eV for most but were 0.2 eV for VN$_6$ in
  afmI Fe and VN$_5$ in Ni and 0.3 eV for VN$_6$ in Ni.}
\end{table}

\subsubsection{V$_m$He$_n$ clusters}

\begin{figure*}
\subfigure[\ afmD Fe, He binding]{\label{DeltaEbHeafmDFig}\includegraphics[width=\columnwidth]{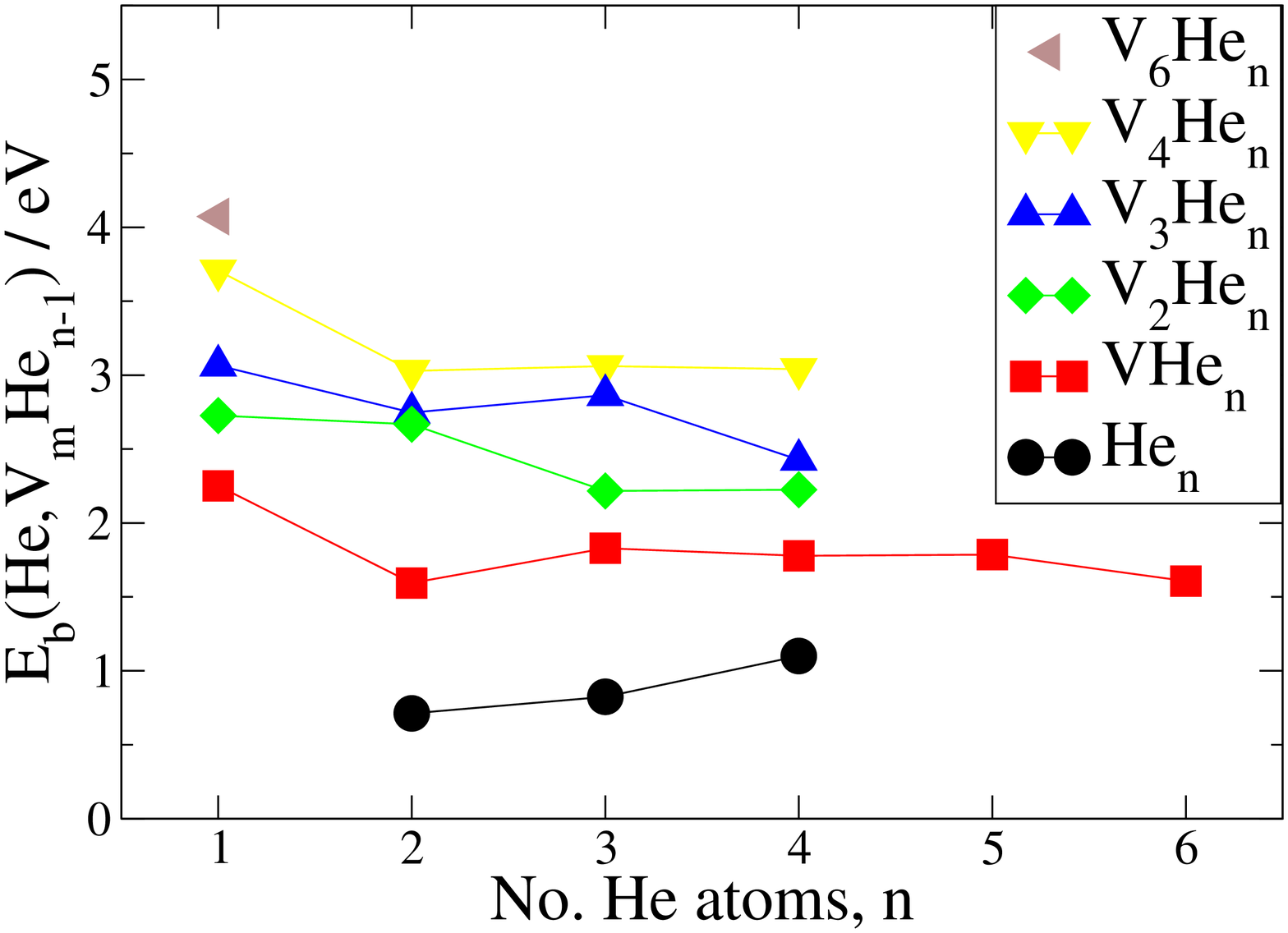}}
\subfigure[\ afmI Fe, He binding]{\label{DeltaEbHeafmIFig}\includegraphics[width=\columnwidth]{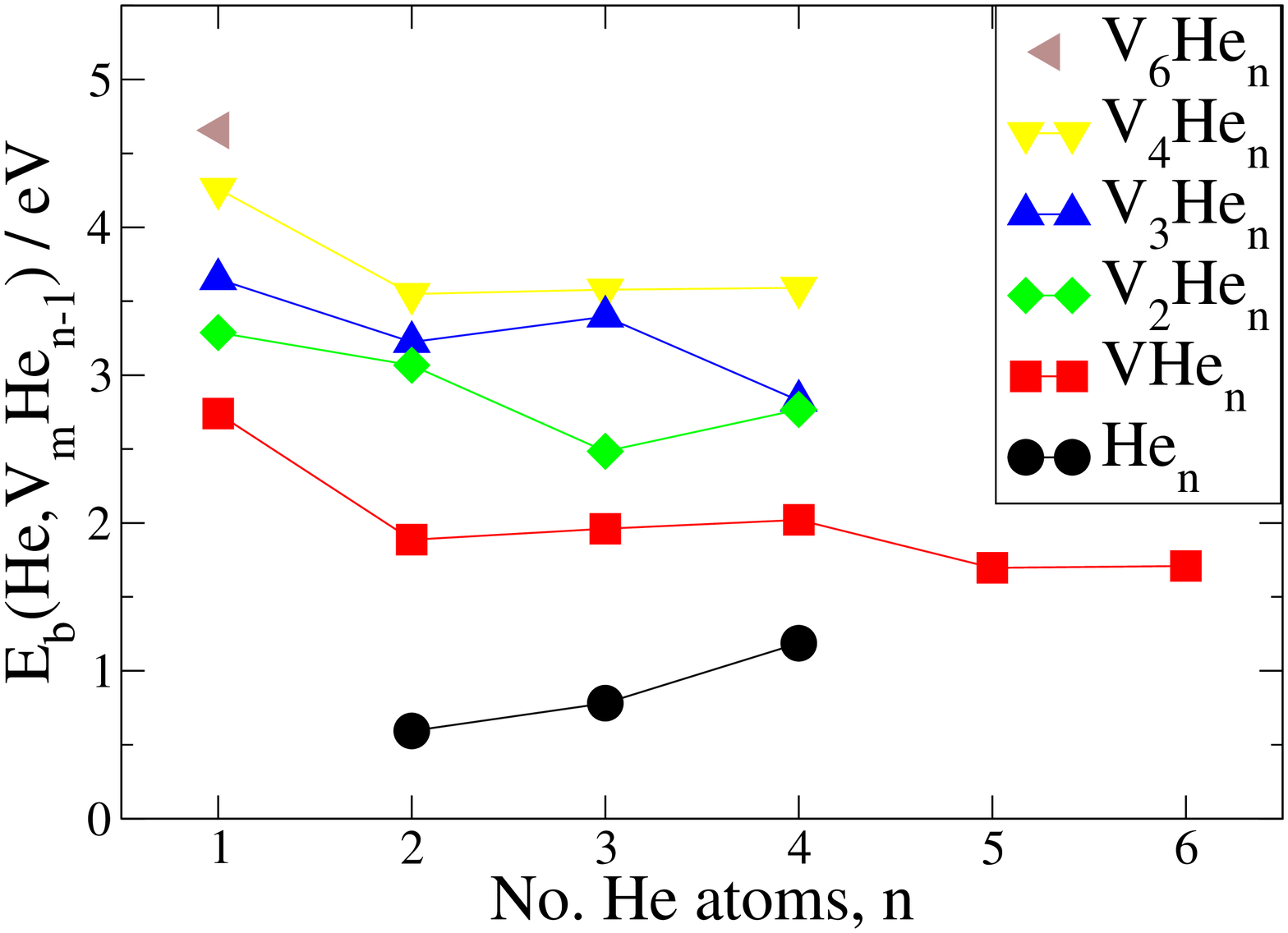}}
\subfigure[\ afmD Fe, V binding]{\label{DeltaEbVafmDFig}\includegraphics[width=\columnwidth]{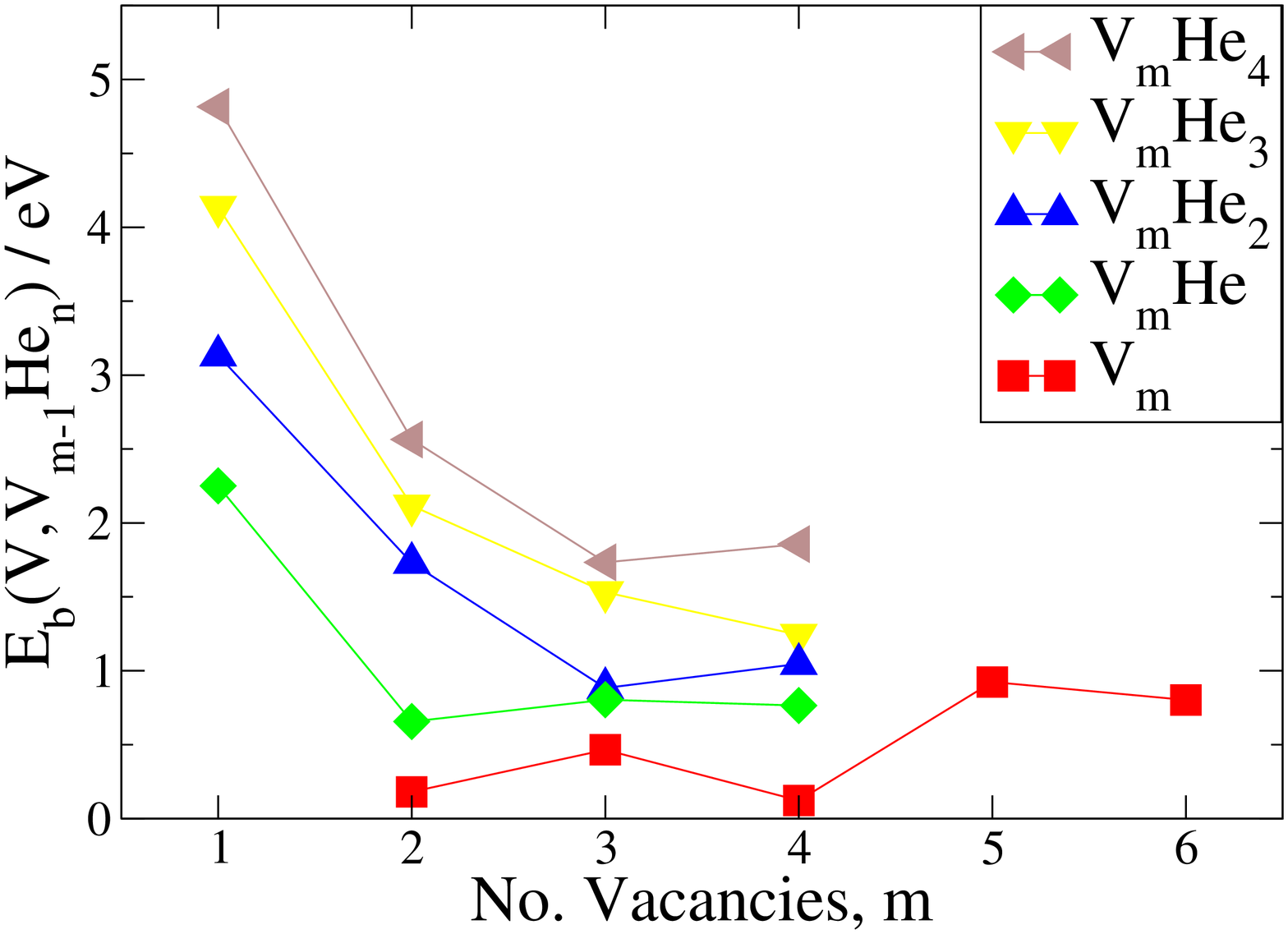}}
\subfigure[\ afmI Fe, V binding]{\label{DeltaEbVafmIFig}\includegraphics[width=\columnwidth]{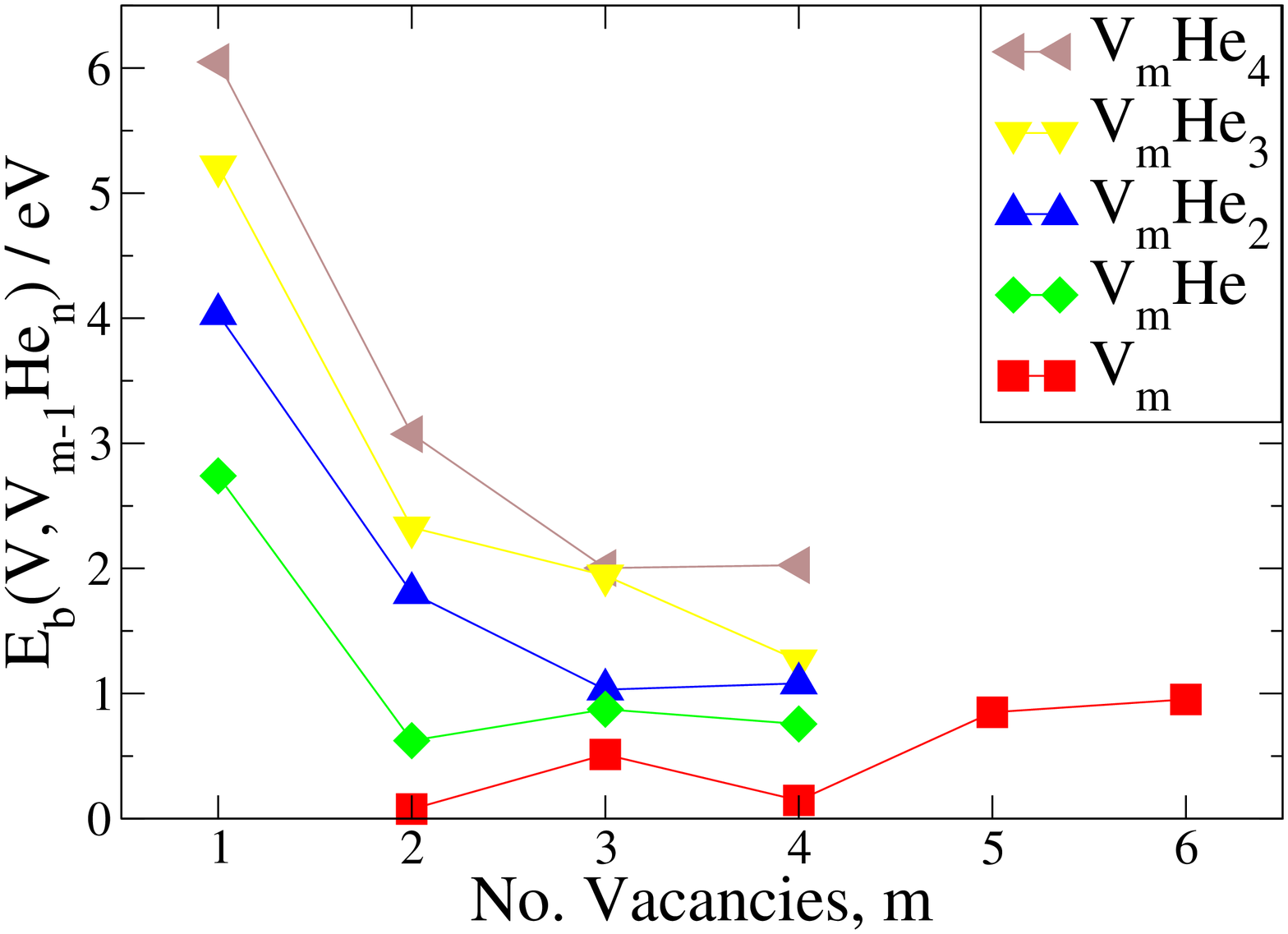}}
\subfigure[\ afmD Fe, SI binding]{\label{DeltaEbSIafmDFig}\includegraphics[width=\columnwidth]{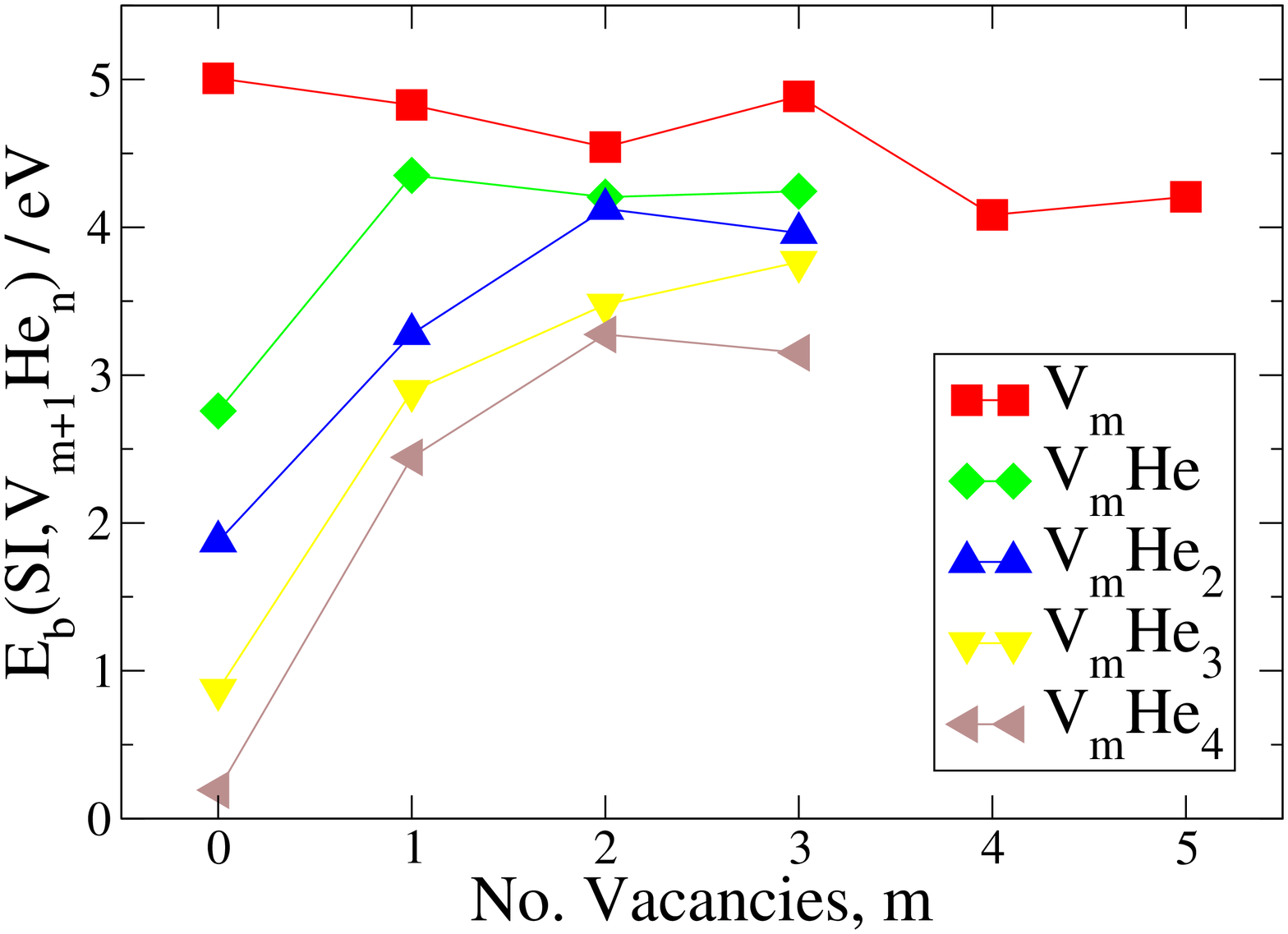}}
\subfigure[\ afmI Fe, SI binding]{\label{DeltaEbSIafmIFig}\includegraphics[width=\columnwidth]{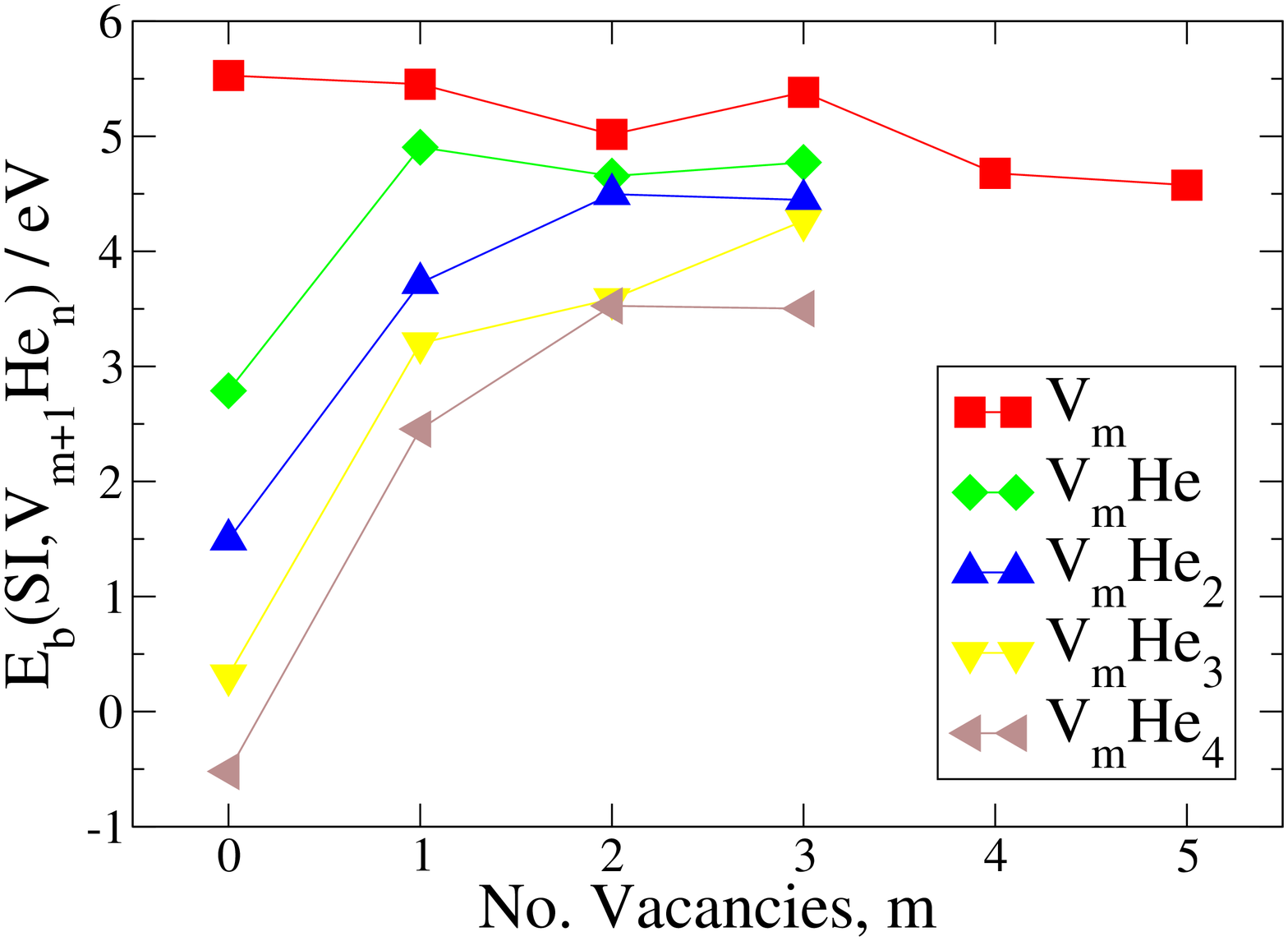}}
\caption{\label{VmHenFig} Binding energies, in eV, for a He atom, V or
  SI to an existing cluster to form one with the V$_m$He$_n$
  stoichiometry in afmD Fe (figures (a), (c) and (e)) and afmI Fe
  (figures (b), (d) and (f)) Fe. Interstitial He cluster data has been
  included in figures (a) and (b) for completeness.}
\end{figure*}

The geometries of the relaxed V$_m$He$_n$ clusters were constrained by
the tendency to maximise He-He and He-Fe separations within the
available volume and, therefore, minimise the repulsive
interactions. In a single vacancy we found that this led to the
following structures: two He formed dumbbells centred on the vacancy
with He-He bond lengths around 1.5 $\angs$; three He formed a
near-equilateral triangle with bond lengths of between 1.6 and 1.7
$\angs$; four He formed a near-regular tetrahedron with bond lengths
between 1.6 and 1.7 $\angs$; five He formed a near-regular triangular
bipyramid with bond lengths between 1.6 and 1.8 $\angs$ and six He
formed a near-regular octahedron with bond lengths between 1.6 and 1.8
$\angs$. In clusters with more than one vacancy, a single He atom
relaxed to a central position. Additional He tended to form similar
clusters to those seen in a single vacancy but now around the centre
of the vacancy cluster. The tri-vacancy case is interesting because
previous DFT calculations in austenite\cite{KlaverFeNiCr} found that a
configuration consisting of a tetrahedral arrangement of vacancies
with one Fe atom near the centre of the void, which can be considered
as the smallest possible stacking fault tetrahedron (SFT), was more
stable than the planar defect of three vacancies with mutual 1nn
separations. The addition of a single He atom was enough to reverse
the order of stability with a difference in the total binding energy
of 0.8 eV in afmI Fe, in favour of the planar defect. We suggest that
this result should readily generalise, with planar defects being more
stable than SFTs with sufficient addition of He. That said, however,
planar defects have been found\cite{KlaverFeNiCr} to be less stable
than three-dimensional proto-voids and this situation is unlikely to
change with the addition of He due to the greater free volume of the
latter clusters.

In Section \ref{HeSoluteSection} the addition of a single He to a
vacancy was found to have very little effect on the local
magnetism. The addition of He to vacancy clusters was generally found
to have very little effect on the total magnetic moment of the
supercells containing the cluster. The only exception was for the
single vacancy in afmD Fe, although it took the addition of six He
atoms to significantly change the magnetic moment. Even in the absence
of vacancies, a cluster of at least three He atoms was necessary to
influence the total magnetic moment. 

In \reffig{VmHenFig} we present results for the binding energy of
either a He atom, vacancy (V) or [001] self-interstitial dumbbell (SI)
to an already existing cluster to form one with the V$_m$He$_n$
stoichiometry. These results show that He consistently binds strongly
to an existing cluster and that the strength of the binding only
increases with $m$. For a fixed value of $n$, this binding energy will
converge to the formation energy for interstitial He [see
  \reftab{solutesInIronTab}] as $m$ increases and is well on the way
to doing that for $n=1$. For fixed $m$ the additional He binding
energy appears to plateau as $n$ increases although it should diminish
eventually as the pressure within the cluster builds.

The binding energy for an additional vacancy is also, consistently,
positive. The presence of He significantly increases this additional
binding for all values of $m$, which is consistent with the
observation that it aids the nucleation, stabilisation and growth of
voids in irradiated
environments\cite{Mansur86,Murase98,Liu04,vanVeen03,Lei11}. For fixed
$n$, the data shows that the vacancy binding energy is tending to a
plateau as $m$ increases and is consistent with the fact that all of
these curves should converge to the vacancy formation energy. 

The SI binding energy can be related to the vacancy binding
energy as follows,
\be
   \ebind(\mathrm{SI},\mathrm{V}_{m+1}\mathrm{He}_n) = \eform(\mathrm{SI}) + \eform(\mathrm{V}) - \ebind(V,\mathrm{V}_m\mathrm{He}_n),
\ee

which implies that the spontaneous emission of an SI from an existing
cluster will be energetically favourable if and only if the binding of
the newly created vacancy is greater than the Frenkel pair formation
energy. The data shows that the SI binding energy clearly decreases as
He concentration is increased at fixed $m$ and for sufficiently high
concentration will become negative. Indeed, it is energetically
favourable for an interstitial He cluster with four He atoms in afmI
Fe, and most likely for five He atoms in afmD Fe, to spontaneously
emit an SI defect. This mechanism was proposed to explain the
observation of He bubbles in Au samples after sub-threshold He
implantation\cite{Thomas81} and could also explain observations of He
trapping in Ni\cite{Thomas79}, where the He was introduced by natural
tritium decay to avoid implantation-produced defects. Our results show
that this would, most likely, occur in austenite and austenitic alloys
and could lead to bubble formation, with the potential for blistering
in the presence of, even low-energy, bombardment by He ions, as seen
in W\cite{Nicholson78,Yoshida81}.

As a whole, the binding energy data is qualitatively similar to DFT
results in Al\cite{Yang} and Pd\cite{Zeng09} and is quantitatively
similar to results in bcc Fe\cite{Fu0507} and
Ni\cite{DomainPerfect,Perfect}. This observation gives us confidence
that our results are, not only, applicable to austenite but to
austenitic alloys more generally.

\begin{figure*}
\subfigure[\ afmD Fe]{\label{EdissafmDFig}\includegraphics[width=\columnwidth]{Fig12a_VmHen_Ediss_vs_ratio_afmD.eps}}
\subfigure[\ afmI Fe]{\label{EdissafmIFig}\includegraphics[width=\columnwidth]{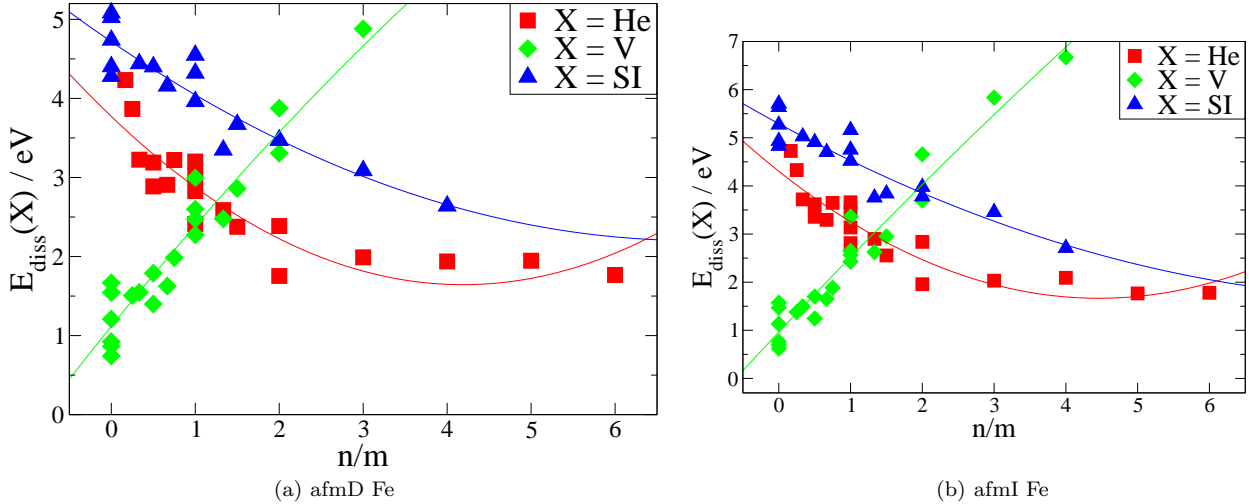}}
\caption{\label{EdissFig} Dissociation energies,
  $E_\mathrm{diss.}(\mathrm{X})$, in eV, for species X from a
  V$_m$He$_n$ cluster, where X is a He, V or SI. Results are presented
  for (a) afmD Fe and (b) afmI Fe versus the He to vacancy ratio,
  $n/m$. The solid curves are simple polynomial fits to the data and
  are present to aid visualisation.}
\end{figure*}

The binding energy data above has also been used to determine the
dissociation energy, that is the energy of emission, of He, V or SI
from a V$_m$He$_n$ cluster using the simple ansatz that the
dissociation energy, $E_\mathrm{diss.}(\mathrm{X})$, for species, X,
is given by
\be
   E_\mathrm{diss.}(\mathrm{X}) = \ebind(\mathrm{X}) + \emig(\mathrm{X}),
\ee
where $\emig(\mathrm{X})$ is the migration energy for isolated
species, X. We present results for the dissociation energies in
\reffig{EdissFig}, using the migration energies in \reftab{EmigTab}.

\begin{table}[htbp]
\begin{ruledtabular}
\begin{tabular}{ccc}
Species,X & afmD Fe & afmI Fe \\
\hline
He & 0.160 & 0.070 \\
V & 0.743 & 0.622 \\
SI & 0.196 & 0.254 \\
\end{tabular}
\end{ruledtabular}
\caption{\label{EmigTab} Migration energies, $\emig(\mathrm{X})$, in
  eV, for Species, X, where X is He, V or SI. For He, the lowest
  values from \reftab{IntHeMigTab} were used. For V, the lowest
  vacancy migration energies from Klaver {\it et
    al.}\cite{KlaverFeNiCr} were used. The SI migration energies were
  calculated here as that for a [100] dumbbell SI migrating between
  two lattice sites at 1nn separation within a magnetic plane using
  identical settings to Klaver {\it et al.}\cite{KlaverFeNiCr}}.
\end{table}

There is a strong and distinct dependence on the He to vacancy ratio,
$n/m$, for the dissociation energies of the three species. Both graphs
exhibit a clear crossover between the He and V curves at around
$n/m=1.3$ and another between the He and SI curves at about
$n/m=6$. An identical He-V crossover ratio was found in bcc
Fe\cite{Fu0507} and fcc Al\cite{Yang}. For $n/m$ below 1.3 the
clusters are most prone to emission of a vacancy, between 1.3 and 6 He
has the lowest dissociation energy and above 6 SI emission is the
preferred dissociation product. The slope of the curves ensures that
emission of the species with the lowest dissociation energy will make
the resulting cluster more stable. At sufficiently high temperatures
that these processes are not limited by kinetics this should lead to
the formation of the most stable clusters, which have an $n/m$ value
at the He-V crossover, where our results predict a minimum
dissociation energy of around 2.8 eV in both afmD and afmI Fe.

\subsubsection{V$_m$C$_n$ and V$_m$N$_n$ clusters}

\begin{table}[htbp]
\begin{ruledtabular}
\begin{tabular}{ccccccc}
 & \multicolumn{2}{c}{afmD Fe} & \multicolumn{2}{c}{afmI Fe} & \multicolumn{2}{c}{Ni} \\
Config. & $\eform$ & $\ebind$ & $\eform$ & $\ebind$ & $\eform$ & $\ebind$ \\
\hline
\multicolumn{7}{c}{VC$_2$ clusters} \\
\hline
1a-1a (opp.) & -16.226 & 0.444 & -16.551 & 0.795 & -15.692 & 0.199 \\
1b-1c & -16.166 & 0.385 & -16.466 & 0.711 & \multicolumn{2}{c}{as 1a-1a (opp.)} \\
1a-1a (adj.) & -15.908 & 0.127 & -16.163 & 0.408 & -15.396 & -0.097 \\
1a-1b & -15.832 & 0.050 & -16.116 & 0.361 & \multicolumn{2}{c}{as 1a-1a (adj.)} \\
1a-1c & -16.104 & 0.323 & \multicolumn{2}{c}{as 1a-1b} & \multicolumn{2}{c}{as 1a-1a (adj.)} \\
$[100]$ dumb. & -16.087 & 0.305 & -16.224 & 0.469 & -15.915 & 0.422 \\
$[001]$ dumb. & -16.265 & 0.484 & -16.262 & 0.507 & \multicolumn{2}{c}{as $[100]$ dumb.} \\
$[110]$ dumb. & -15.407 & -0.375 & -15.238 & -0.517 & -15.380 & -0.113 \\
$[111]$ dumb. & \multicolumn{4}{c}{rlx $[001]$ dumb.} & -15.565 & 0.072 \\
\hline
\multicolumn{7}{c}{VN$_2$ clusters} \\
\hline
1a-1a (opp.) & -16.373 & 0.981 & -16.579 & 1.295 & -14.559 & 0.872 \\
1b-1c & -16.124 & 0.732 & -15.932 & 0.648 & \multicolumn{2}{c}{as 1a-1a (opp.)} \\
1a-1a (adj.) & -16.113 & 0.720 & -16.237 & 0.953 & -14.279 & 0.591 \\
1a-1b & -16.051 & 0.658 & -16.037 & 0.753 & \multicolumn{2}{c}{as 1a-1a (adj.)} \\
1a-1c & -16.228 & 0.835 & \multicolumn{2}{c}{as 1a-1b} & \multicolumn{2}{c}{as 1a-1a (adj.)} \\
$[100]$ dumb. & -14.451 & -0.942 & -14.487 & -0.796 & -13.616 & -0.071 \\
\end{tabular}
\end{ruledtabular}
\caption{\label{vactwosolTab} Formation and total binding energies, in
  eV, for the interactions of a vacancy with two octa sited C or N
  solutes. Configurations with C or N in octa sites at 1nn to the
  vacancy are labelled by the positions of the two solutes as in
  \reffig{SubOctaFig}. When both octa solutes are in the same plane as
  the vacancy the configurations are additionally labelled by their
  relative orientation i.e. opposite (opp.) or adjacent (adj.)  to one
  another. Doubly mixed dumbbells centred on the vacancy site were
  also considered as configurations of an interacting vacancy with two
  octa solutes and the total binding energies were calculated
  accordingly. Eshelby corrections to both $\eform$ and $\ebind$ were
  found to be no more than 0.03 eV in magnitude.}
\end{table}

In fct afmD and afmI Fe and in Ni we considered VX$_n$ clusters with
octa-sited C and N at 1nn to the vacancy and configurations where C
and N are close enough to form C-C and N-N bonds within the
vacancy. Our results for VC$_2$ and VN$_2$ clusters are given in
\reftab{vactwosolTab}. We found that VX$_2$ clusters with octa-sited C
and N are most stable when the C/N atoms are as far apart as possible,
that is, opposite one another across the vacancy. For these
configurations the total binding energy is more than the sum of the
binding energies for each single solute to the vacancy, indicating
either some chemical or cooperative strain interaction. We found that
C-C dumbbells centred on the vacancy are stable, with bond lengths
between 1.38 and 1.48 $\angs$, that is, much shorter than the
separations between octahedral sites. The most stable lie along
$\langle 100\rangle$ directions and binding over and above that for
octa sited C was found in afmD Fe and Ni. The enhancement in binding
upon forming a C-C dumbbell is not, however, as pronounced as was seen
in bcc Fe\cite{Domain04,Fu0507,Forst06,Lau07}. We also found stable
configurations with N-N dumbbells in Fe and Ni with bond lengths
between 1.34 and 1.49 $\angs$, although they exhibit a much lower, and
generally negative, total binding energy compared to configurations
with octa sited N atoms.

For VX$_3$ clusters, we investigated all possible configurations with
three octa-sited C or N solutes in addition to those with a C-C
dumbbell and an octa-sited C solute in one of the four octa sites
perpendicular to the dumbbell axis and a configuration with three C
atoms close enough for C-C bonding. Although we do find stable
configurations with C-C bonding in either a dumbbell or triangular
arrangement in both Fe and Ni, these arrangements are the least stable
and exhibit significant, negative total binding energies. The most
stable arrangements consist of three octa sited C atoms placed as far
apart as possible, for example in three 1a sites relative to the
vacancy as in \reffig{SubOctaFig}. However, the total binding energies
for the most stable VC$_3$ clusters [see \reftab{clusterBindingTab}]
are less than for VC$_2$, which implies that a vacancy can only bind
up to two C atoms within a vacancy. A vacancy may still, however, bind
more than at 2nn octa sites but we did not investigate this
possibility due to the strongest binding being at 1nn to the vacancy
and due to the large number of possible configurations.

The most stable VN$_3$ clusters have the same geometry as found with C
but, in contrast, are more stable than VN$_2$ clusters. Beyond this
point, we found that the total binding energy only increases for up to
four N atoms in afmI Fe and Ni but increases all the way up to six N
atoms in afmD Fe. That said, however, the binding energy per N atom
only increases up to a VN$_2$ cluster in all reference states. The
equilibrium concentrations of clusters with more than two N atoms,
which can be calculated using the law of mass-action\cite{Fu08}, would
very likely be negligible, even at room temperature. Despite their
magnitude, the Eshelby corrections do not change these conclusions but
would result in the total binding energy increasing all the way up to
six N atoms in afmI Fe, as was found for afmD Fe.

We investigated site preference and binding for single C and N solutes
to the most stable di-, tetra- and hexa-vacancy clusters in afmD and
afmI Fe and in Ni. For the V$_2$C cluster in afmD Fe, we considered
all 1nn octa sites to the three distinct types of 1nn di-vacancy as
well as configurations with C at the centre of all 1nn and 2nn
di-vacancy clusters. For the octa sites, C was found to bind to
existing di-vacancy clusters with similar binding energies to a single
vacancy, that is with $\ebind(\mathrm{C},\mathrm{V}_2)$ in the range
from 0.03 to 0.32 eV. The most stable of these, which was also found
to be the most stable V$_2$C cluster, contained the most stable
di-vacancy and bound C more stably than to a single vacancy. We found
that C was repelled from the centre of a 1nn di-vacancy lying within a
magnetic plane but bound to the other two 1nn di-vacancies with
similar energies to those found in octa sites. As in bcc
Fe\cite{Forst06,Lau07,Ortiz0709}, the most preferred site for C was at
the centre of a 2nn di-vacancy, with $\ebind(\mathrm{C},\mathrm{V}_2)
= 0.35$ eV. However, this was not sufficient to overcome the
difference in stability between 1nn and 2nn di-vacancies in afmD
Fe\cite{KlaverFeNiCr} and did not, therefore, form them most stable
V$_2$C cluster, in contrast to in bcc Fe.

The analysis above motivated the use of only the most stable 1nn
di-vacancy in the remaining calculations along with configurations
containing solutes at the centre of 2nn di-vacancies. For N in afmD
Fe, the order of site preference mirrors that for C. An N solute is
capable of stabilising a 2nn di-vacancy configuration but the most
stable V$_2$N cluster shared the same geometry as for C with a binding
energy to the underlying di-vacancy of 0.56 eV, which is, again, in
excess of the binding to a single vacancy.

The situation in afmI Fe and Ni was found to be rather similar to that
of afmD Fe. For both C and N, the site at the centre of a 1nn
(in-plane) di-vacancy was disfavoured. The most stable configuration
generally contained an octa-sited solute bound to a 1nn
di-vacancy. The only exception was in afmI Fe, where a configuration
with C at the centre of a 2nn di-vacancy lying within a magnetic plane
had a greater total binding energy but only by 0.03 eV. This, most
likely resulted from the much smaller energy difference between 1nn
and 2nn di-vacancies in afmI Fe of compared to afmD
Fe\cite{KlaverFeNiCr} and to Ni, where we find an energy difference of
0.1 eV in favour of the 1nn di-vacancy. In the most stable clusters,
the binding of the solutes to the underlying di-vacancy was, once
again, in excess of the binding to a single vacancy.

For the binding of C and N to the most stable tetra-vacancy, we found
that the central position was extremely disfavoured. We investigated
all configurations with solute atoms in an octa site at 1nn to at
least a single vacancy. We also performed calculations with solute
atoms placed initially at random within the proto-void but found that
these relaxed to octa sites already considered. Configurations with
only a single vacancy at 1nn to the solute were found to be the most
stable. The total binding energies for these configurations are given
in \reftab{clusterBindingTab}. Using these results we found that the
binding of C and N to the tetra-vacancy was in excess of that to a
di-vacancy and a single vacancy, in all cases except for N in afmI Fe,
where the binding to the tetra-vacancy and di-vacancy reversed order,
although they only differ by 0.02 eV.

For the hexa-vacancy, the central octa site was unstable for both C
and N in afmD Fe. In afmI Fe and Ni it was stabilised by symmetry but
still strongly disfavoured. This repulsion is, however, significantly
less than was observed for the tetra-vacancy. Closer observation
showed that while the nearest neighbouring Fe and Ni atoms to the
solutes moved very little under relaxation in the tetra-vacancy, the
contraction in bond length was between 25 and 30 \% in the
hexa-vacancy from an initial separation of around 3 $\angs$. This
demonstrates how important the formation of strong chemical bonds with
characteristic bond lengths is to the stability of configurations
containing C and N in Fe and Ni.

We investigated the stability of configurations with C and N in all
octa sites at 1nn to at least one vacancy in the hexa-vacancy
cluster. We found that there were additional stable sites, lying along
$\langle 100\rangle$ axes projected out from the centre of the
hexa-vacancy. For C, these sites were found to lie between the first
vacancy reached along these axes and the next octa site out. They are
close to but distinct from the octa sites and we, therefore, refer to
them as octa-b sites. For N, stable sites were found between the
centre of the hexa-vacancy and the first vacancy reached along the
$\langle 100\rangle$ axes and we refer to these as off-centre
sites. We found that C was, consistently, most stable in an octa-b
site whereas N preferred octa sites with two vacancies at 1nn,
although an off-centre site along $[00\bar{1}]$ was the most stable in
afmD Fe.

\begin{figure}
\includegraphics[width=\columnwidth]{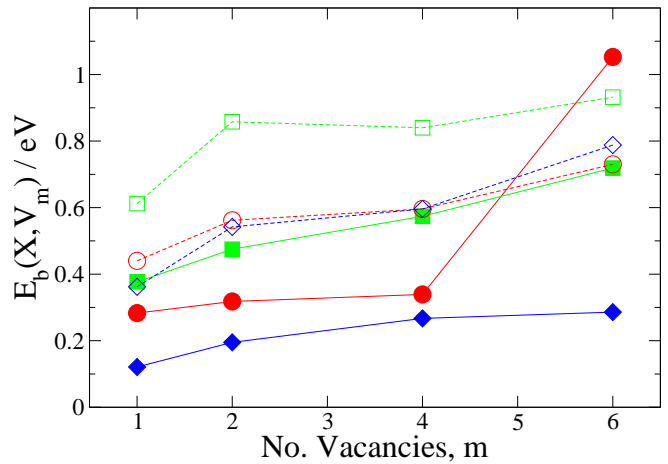}
\caption{\label{XVmFig} Binding energy, $\ebind(\mathrm{X},\mathrm{V}_m)$, in eV, where X is C (filled symbols and solid lines) or N (unfilled symbols and dashed lines) in afmD Fe (red circles), afmI Fe (green squares) and Ni (blue diamonds). The binding energies were calculated for the most stable clusters.}
\end{figure}

Once again, the binding energy between the solutes and hexa-vacancy
was greater than for all smaller vacancy clusters. We summarise these
results for $\ebind(\mathrm{X},\mathrm{V}_m)$ in \reffig{XVmFig},
which clearly shows the increase in binding energy as the vacancy
cluster becomes larger. It also clearly shows that in the same
reference state, the binding energy for N is consistently greater than
for C and that the binding energies in afmI Fe lie above those in afmD
Fe. The one anomalous point is the binding energy for C to a
hexa-vacancy in afmD Fe, which is much larger than the trends would
suggest. Other configurations with C in an octa-b site in afmD Fe
exhibited similar levels of binding and no problems with any of these
calculations or instabilities in the relaxed structures could be
found.

\subsection{$[001]$ dumbbell SI - solute interactions}

We investigated the binding of He, C and N solutes to a [001] dumbbell
in afmD and afmI Fe and in Ni and present the results in
\reftab{sisolTab}.

\begin{table}[htbp]
\begin{ruledtabular}
\begin{tabular}{ccccccc}
 & \multicolumn{2}{c}{afmD Fe} & \multicolumn{2}{c}{afmI Fe} & \multicolumn{2}{c}{Ni} \\
A-B/Config. & $\eform$ & $\ebind$ & $\eform$ & $\ebind$ & $\eform$ & $\ebind$ \\
\hline
$[001]$-tetra He/1a & \multicolumn{2}{c}{unstable} & \multicolumn{2}{c}{unstable} & & \\
$[001]$-tetra He/1b & 7.734 & -0.075 & \multicolumn{2}{c}{as 1a} & & \\
$[001]$-tetra He/2a & 7.494 & 0.166 &  8.517 & 0.098 & & \\
$[001]$-tetra He/2b & 7.761 & -0.036 & \multicolumn{2}{c}{as 2a} & & \\
$[001]$-tetra He/2c & 7.609 & 0.117 & 8.435 & 0.180 & & \\
$[001]$-tetra He/2d & 7.514 & 0.146 & \multicolumn{2}{c}{as 2c} & & \\
\hline
$[001]$-sub He/1a & 7.045 & 0.176 & 7.743 & 0.089 & 7.085 & 0.235 \\
$[001]$-sub He/2a & 7.164 & 0.057 & 7.815 & 0.017 & 7.316 & 0.003 \\
$[001]$-sub He/2b & 7.050 & 0.171 & 7.653 & 0.179 & \multicolumn{2}{c}{unstable} \\
$[001]$-sub He/2c & 7.095 & 0.126 & \multicolumn{2}{c}{as 2b} & \multicolumn{2}{c}{as 2b} \\
\hline
$[001]$-C/1a & -5.563 & -0.037 & -5.007 & -0.202 & -4.300 & 0.012 \\
$[001]$-C/1b & -4.604 & -0.997 & -3.975 & -1.234 & & \\
$[001]$-C/1c & -4.459 & -1.141 & \multicolumn{2}{c}{as 1b} & & \\
$[001]$-C/2a & -5.585 & -0.015 & -5.064 & -0.145 & -4.322 & 0.034 \\
$[001]$-C/2b & -5.527 & -0.074 & \multicolumn{2}{c}{as 2a} & \multicolumn{2}{c}{as 2a} \\
$[001]$-C/4a & -5.626 & 0.025 & -5.266 & 0.057 & -4.303 & 0.015 \\
$[001]$-C/4b & -5.652 & 0.051 & -5.287 & 0.078 & -4.362 & 0.075 \\
$[001]$-C/4c & -5.642 & 0.041 & \multicolumn{2}{c}{as 4b} & \multicolumn{2}{c}{as 4b} \\
\hline
$[001]$-N/1a & -5.106 & -0.300 & -4.444 & -0.529 & -3.198 & -0.188 \\
$[001]$-N/1b & -4.461 & -0.945 & -3.721 & -1.252 & & \\
$[001]$-N/1c & -4.197 & -1.209 & \multicolumn{2}{c}{as 1b} & & \\
$[001]$-N/2a & -5.290 & -0.116 & -4.762 & -0.211 & -3.400 & 0.015 \\
$[001]$-N/2b & -5.251 & -0.155 & \multicolumn{2}{c}{as 2a} & \multicolumn{2}{c}{as 2a} \\
$[001]$-N/4a & -5.425 & 0.019 & -5.027 & 0.054 & 3.402 & 0.017 \\
$[001]$-N/4b & -5.430 & 0.024 & -5.043 & 0.069 & -3.481 & 0.096 \\
$[001]$-N/4c & -5.458 & 0.052 & \multicolumn{2}{c}{as 4b} & \multicolumn{2}{c}{as 4b} \\
\end{tabular}
\end{ruledtabular}
\caption{\label{sisolTab} Formation and binding energies in eV for
  [001] self-interstitial dumbbell (species A) - solute (species B)
  interactions. For octa-sited C and N solutes the configurations are
  labelled as in \reffig{SubOctaFig}. He interactions were
  investigated with He sited substitutionally and tetrahedrally with
  configurations labelled as in \reffigs{SubSubFig}{SubTetraFig},
  respectively. Configurations with substitutional He in 1b and 1c
  positions relative to a [001] dumbbell SI were unstable to defect
  recombination and interstitial He kick-out. In Ni, the [001]-tetra
  He binding energy was observed to be 0.20
  eV\cite{DomainPerfect}. Eshelby corrections were found down to -0.1
  eV for configurations containing substitutional He but the related
  increases in $\ebind$ were no more than 0.02 eV. For configurations
  containing interstitial solutes $E^\mathrm{corr.}$ could be as low
  as -0.2 eV with corresponding increases in $\ebind$ up to 0.08 eV.}
\end{table}

We found that interstitial He, placed initially 1nn to a [001] SI
dumbbell, either spontaneously displaced under relaxation to a 2nn
site or exhibited a repulsive binding energy in Fe. At 2nn, however, a
positive binding energy was observed, up to almost 0.2 eV, as was
found in Ni\cite{DomainPerfect,Perfect}. Eshelby corrections do not
qualitatively change these results and would only act to enhance the
binding at 2nn. This positive binding energy is comparable to that in
bcc Fe\cite{Fu0507} but while significant, it is only likely to result
in mutual trapping at low temperature, given the high mobility of the
two species. Taken as a model for the binding of interstitial He to
other overcoordinated defect sites, such as near dislocations and
grain boundaries, however, this result does show that He would be
likely to be trapped at such sites, leading to interstitial He cluster
formation and spontaneous bubble nucleation and growth, as discussed
earlier. It is worth mentioning that bubble nucleation by this
mechanism would happen much more readily at grain boundaries where,
due to their disorder, vacancies can be formed without the additional
SI.

A substitutional He atom in the 1b and 1c sites [see
  \reffig{SubSubFig}] to a [001] SI dumbbell resulted in the
spontaneous recombination of the vacancy and SI and the kick-out of an
interstitial He atom. At all other 1nn and 2nn sites except 2b in Ni,
however, stable complexes with binding energies of up to around 0.2 eV
were formed. Barriers to recombination for these complexes, while
positive, were not calculated in this work. These results do, however,
show that substitutional He, and most likely, other V$_m$He$_n$
clusters can act as trapping sites for SI dumbbells in austenite and
austenitic alloys with a capture radius extending out to at least
2nn. We can also speculate that, once trapped, recombination will be
likely to occur.

Both C and N are either repelled from 1nn and 2nn sites to a [001] SI
dumbbell or show very little positive binding, much as was observed in
bcc Fe\cite{Domain04}. Eshelby corrections do not change this
conclusion in Fe but would result in binding of around 0.1 eV at 2nn
in Ni. Motivated by the result that C does exhibit positive binding to
the most stable SI and small SI clusters in bcc Fe\cite{Perfect} at
further separation, we investigated this possibility here and found
sites with binding energies from 0.05 to 0.1 eV at 4nn to the
dumbbell, which would only be enhanced by Eshelby corrections. These
sites can be related to the corresponding ones in bcc Fe by a Bain
transformation\cite{Bain} and the binding almost certainly results
from strain field effects in both cases. The fact that such binding
was found to increase with interstitial cluster size\cite{Perfect}
means that Cottrell atmospheres\cite{Cottrell49} of C and N are very
likely to form around other overcoordinated defects, such as
dislocations and grain boundaries, in both ferritic and austenitic
alloys under conditions where these species are mobile.

\section{Conclusions}

An extensive set of first-principles DFT calculations have been
performed to investigate the behaviour and interactions of He, C an N
solutes in austenite, dilute Fe-Cr-Ni alloys and Ni as model systems
for austenitic steel alloys. In particular, we have investigated the
site stability and migration of single He, C and N solutes, their
self-interactions, interactions with substitutional Ni and Cr solutes
and their interactions with point defects typical of irradiated
environments, paying particular attention to the formation of small
V$_m$X$_n$ clusters.

Direct comparison with experiment verifies that the two-state approach
used to model austenite in this work is reasonably
predictive. Overall, our results demonstrate that austenite behaves
much like other fcc metals and is qualitatively similar to Ni in many
respects. We also observe a strong similarity between the results
presented here for austenite and those found previously for bcc Fe.

We find that interstitial He is most stable in the tetrahedral site
and migrates via off-centre octahedral transition states with a
migration energy from 0.1 to 0.2 eV in austenite and 0.13 eV in
Ni. The similarity of these results and the weak interactions with Ni
and Cr solutes in austenite suggests a migration energy in Fe-Cr-Ni
austenitic alloys in the 0.1 to 0.2 eV range. Interstitial He will,
therefore, migrate rapidly from well below room temperature until
traps are encountered. Its strong clustering tendency, with an
additional binding energy approaching 1 eV per He atom in austenite
and 0.7 eV in Ni, will lead to a reduction in mobility as interstitial
He concentration increases. Interactions with overcoordinated defects,
which are on the order of a few tenths of 1 eV, will result in the
build up and clustering of interstitial He at dislocations and grain
boundaries. The most stable traps, however, are vacancy clusters and
voids, with binding energies of a few eV. The strength of this binding
means that growing interstitial He clusters eventually become unstable
to spontaneous Frenkel pair formation, resulting in the emission of a
self-interstitial and nucleation of a VHe$_n$ cluster. The binding of
additional He and vacancies to existing V$_m$He$_n$ clusters increases
significantly with cluster size, leading to unbounded growth and He
bubble formation in the presence of He and vacancy fluxes. The most
stable clusters have a helium to vacancy ratio, $n/m$, of around 1.3,
with a dissociation energy for the emission of He and V of 2.8 eV in
austenite and Ni. Generally, we assume that V$_m$He$_n$ clusters are
immobile. For the simplest case of substitutional He, however,
migration is still possible. In a thermal vacancy population,
diffusion by the dissociative mechanism dominates, with an activation
energy of between 0.6 and 0.9 eV in Fe and 1.4 in Ni. In irradiated
environments, however, the vacancy mechanism dominates and diffusion
can proceed via the formation and migration of the stable V$_2$He
complex, with an activation energy of between 0.3 and 0.6 eV in Fe and
0.8 eV in Ni.

We find that C and N solutes behave similarly, both in austenite and
Ni, although the interactions of N are stronger. The octahedral
lattice site is preferred by both solutes, leading to a net expansion
of the lattice and a reduction of the $c/a$ ratio in the afmD and afmI
Fe reference states. Both solutes also stabilise austenite over
ferrite and favour ferromagnetic over antiferromagnetic states in
austenite. Carbon migrates via a $\langle 110\rangle$ transition state
with a migration energy of at least 1.3 eV in austenite and of 1.6 eV
in Ni. For N, migration proceeds via the crowdion or tetrahedral
sites, depending on path, with a migration energy of at least 1.4 eV
in austenite and 1.3 eV in Ni. Pairs of solute atoms are repelled at
1nn and 2nn in austenite and do not interact in Ni. Both C and N
interact very little with Ni solutes in austenite but bind to Cr,
which may act as a weak trap and encourage the formation of
Cr-carbonitrides in conditions where the solutes are mobile. Carbon
binds to a vacancy by up to 0.4 eV in austenite and 0.1 eV in Ni, with
N binding more strongly at up to 0.6 eV in austenite and 0.4 eV in
Ni. While this may suggest that C and N act as vacancy traps, as in
bcc Fe, preliminary calculations in Ni show that VC and VN clusters
may diffuse cooperatively with an effective migration energy similar
to that for the isolated vacancy. This also raises the possibility of
enhanced C and N mobility in irradiated alloys and their segregation
to defect sinks. A vacancy can bind up to two C atoms and up to six N
atoms in austenite (or four in Ni), although the additional binding
energy reduces significantly above two. Covalent bonding was observed
between solutes in a vacancy but did not lead to any enhanced
stability, as seen in bcc Fe. Both C and N show a strong preference
for sites near the surface of vacancy clusters and the binding
increases with cluster size, suggesting that they will decorate the
surface of voids and gas bubbles, when mobile. A binding energy of 0.1
eV was observed to a $[001]$ SI dumbbell in austenite and Ni, which we
would expect to increase with interstitial cluster size, as in bcc Fe,
resulting in Cottrell atmospheres of C and N around dislocations and
grain boundaries in austenitic alloys.

Along with previous work, these results provide a complete database
that would allow realistic Fe-Cr-Ni austenitic alloy systems to be
modelled using higher-level techniques, such as molecular dynamics
using empirical potentials and kinetic Monte Carlo simulations. As
such, they play a critical role in a multi-scale modelling approach to
study the microstructural evolution of these materials under
irradiation in typical nuclear environments.

\section{Acknowledgements}

This work was part sponsored through the EU-FP7 PERFORM-60 project,
the G8 funded NuFUSE project and EPSRC through the UKCP collaboration.

\end{document}